\documentclass[english]{revtex4-1}
\usepackage[T1]{fontenc}
\usepackage[latin9]{inputenc}
\usepackage[letterpaper]{geometry}
\usepackage{array}
\usepackage{multirow}
\usepackage{amsmath}
\usepackage{amssymb}
\usepackage{graphicx}

\makeatletter

\providecommand{\tabularnewline}{\\}

\usepackage{amsmath}
\usepackage{setspace}

\makeatother

\usepackage{babel}
\begin{document}
\title{Geometric Constraints on Two-electron Reduced Density Matrices}
\author{Yimin Li}
\affiliation{Shcool of Physical Science and Technology, ShanghaiTech University,
Shanghai 201210, China}
\email{liym1@shanghaitech.edu.cn}

\date{10/21/2020}
\begin{abstract}
For many-electron systems, the second-order reduced density matrix
(2-RDM) provides sufficient information for characterizing their properties
of interests in physics and chemistry, ranging from total energy,
magnetism, quantum correlation and entanglement to long-range orders.
Theoretical prediction of the structural properties of 2-RDM is an
essential endeavor in quantum chemistry, condensed matter physics
and, more recently, in quantum computation. Since 1960s, enormous
progresses have been made in developing RDM-based electronic structure
theories and their large-scale computational applications in predicting
molecular structure and mechanical, electrical and optical properties
of various materials. However, for strongly correlated systems, such
as high-temperature superconductors, transition-metal-based biological
catalysts and complex chemical bonds near dissociation limit, accurate
approximation is still out of reach by currently most sophisticated
approaches. This limitation highlights the elusive structural feature
of 2-RDM that determines quantum correlation in many-electron system.
Here, we present a set of constraints on 2-RDM based on the basic
geometric property of Hilbert space and the commutation relations
of operators. Numerical examples are provided to demonstrate the pronounced
violation of these constraints by the variational 2-RDMs. It is shown
that, for a strongly correlated model system, the constraint violation
may be responsible for a considerable portion of the variational error
in ground state energy. Our findings provide new insights into the
structural subtlety of many-electron 2-RDMs. 
\end{abstract}
\maketitle

\section*{Introduction}

In 1930, Paul Dirac introduced the idea of utilizing reduced density
matrix (RDM) to approximate the properties of many-electron systems
in order to avoid intractable computation of many-electron wave function.\citep{Dirac2008}
The research efforts in this direction led to the development of various
electronic structure theories based on one- and two-electron RDMs
(1-RDM and 2-RDM).\citep{Husimi1940,Lowdin1955,Mayer1955,Hohenberg1964,Kohn1965,Parr1989,Levy2005,Mardirossian2017,Verma2020,Coleman1963,Coleman2000,Mazziotti2007,Goedecker2000,Nagy2003,Sharma2008}
Among these approaches, density functional theory (DFT) is most successful
and popular.\citep{Hohenberg1964,Kohn1965,Parr1989,Mardirossian2017}
In DFT, the ground-state energy is approximated by an energy functional
of one-electron density (that is, the diagonal of 1-RDM), which provides
sufficient accuracy with low computational cost for most of quantum
physics and chemistry applications. A major challenge in DFT is to
systematically improve energy functional approximation for describing
strongly correlated systems.\citep{Mardirossian2017,Verma2020,Sun2015,Ekholm2018,Isaacs2018}

In parallel, the approaches based on high-order RDM have been actively
pursued aiming at the systems with strongly correlated electrons or
nuclei.\citep{Husimi1940,Lowdin1955,Mayer1955,Coleman1963,Valdemoro1996,Coleman2000,Mazziotti2007,Mayorga2018}
In these approaches, the energy expression is exact. However, it is
difficult to find sufficient constraints on an approximated RDM in
order to ensure its correspondence to a many-electron wave function.\citep{Coleman2000}
In quantum chemistry, this problem is known as $N$-representability
problem,\citep{Coleman1963} a special case of quantum marginal problem
in quantum information\citep{Klyachko2006} . For strongly correlated
systems, this problem may cause predicting erroneous bond dissociation
barrier, and unphysical properties such as factional charges and factional
spins.\citep{Mihailovic1969,Ayers2007,Cohen2008,Aggelen2009,Nakata_2009,Anderson2013,Ding2020}
Since the problem was formalized in early 1960s, substantial research
efforts have been made to identify sufficiently stringent N-representability
constraints and to implement them in practical computation. The progress
has been steady but slow due to its challenging mathematical and computational
nature.\citep{Coleman1963,Coleman2000,Klyachko2006,Liu2007,Mazziotti2007,Li2018}
With recent exciting development in quantum computing and information
technologies, this problem is attracting more attention because of
the key role of quantum marginals in quantum measurement and information
processing.\citep{McArdle2020,Higuchi2003,Higuchi2003a,Klyachko2004,Liu2007,Cramer2010,Schilling2013,Mazziotti2016a,Rubin2018,Smart2019,Smart2019a,Takeshita2020}

Currently, most of nontrivial $N$-representability constraints are
originated from two basic properties of a state in fermion Fock space:
antisymmetric permutation\citep{Klein1969a,Klein1969,Kryachko1981}
and the fact that inner product of the state with itself is nonnegative.
The symmetry property imposes an upper bound on the eigenvalues of
1-RDM (Pauli principle), and antisymmetric condition on electron and
hole 2-RDM.\citep{Coleman1963} A major breakthrough on the quantum
marginal problem has been made by Klyachko utilizing representation
theory of symmetric group, which leads a family of constraints on
the eigenvalues of pure-state 1-RDM (Generalized Pauli constraints).\citep{Borland1972,Higuchi2003,Higuchi2003a,Klyachko2004,Klyachko2006,Ruskai2007,Klyachko2009,Schilling2013,Schilling2018}
Based on generalized Pauli constraints, pure-state constraints on
2-RDM have been proposed recently.\citep{Mazziotti2016}

The non-negativity inner product property requires any RDM to be Hermitian
and positive semidefinite, which, for 2-RDM, implies a set of constraints
known as: $D$, $Q$, $G$, $T1$, $T2$ and $T2'$ conditions. The
$D$, $Q$ and $G$ conditions were proposed by Coleman\citep{Coleman1963},
Garrod and Percus\citep{Garrod1964} in early 1960s. In 1978, Erdahl
discovered $T1$ and $T2$ conditions by introducing a clever idea
to reduce the 3-RDM positive semidefinite conditions to a set of conditions
on 1-RDM and 2-RDM.\citep{Erdahl1978} The more restrictive $T2'$
condition was reduced from the positive semidefinite condition on
a variant of 3-RDM.\citep{Zhao2004,Hammond2005} Inspired by this
idea, Mazziotti developed a systematic approach to deduce 2-RDM conditions
from higher-order RDM constraints, and proved, in 2012, that inclusion
of the whole set of deduced conditons sufficiently ensure a 2-RDM
to be N-representable.\citep{Mazziotti2012} However, the number of
the deduced conditions increases exponentially with the many-body
order of the RDM. It is not yet clear how the effectiveness of these
conditions depends on the order increase. 

At present, variational 2-RDM method is one of most promising high-order
RDM approaches. In this approach, the positive semidefinite conditions
can be implemented by either positive semidefinite programming (SDP)\citep{Nakata2001}
or nonlinear optimization\citep{Mazziotti2004}. For many molecular
systems with up to 28 electrons, the accuracy of variational ground
state energy is comparable to the high-level wavefunction-based method
CCSD(T).\citep{Zhao2004,Nakata2008} However, for the strongly correlated
systems, such as 1D, quasi-2D and 2D Hubbard models\citep{Hammond2006,Nakata2008,Verstichel_2013,Anderson2013,Rubin2014},
the Lipkin model\citep{Hammond2005}, molecule chains\citep{Mazziotti2004,FossoTande2015,Mazziotti2016a},
and the molecules near dissociation limit\citep{Aggelen2009,Aggelen2010,Nakata2012,Nakata_2009},
the variatonal results are encouraging but still unsatisfactory. Evidently,
more restrictive constraint is in demand to elucidate the intriguing
physical and chemical properties of strong correlated systems.

In this paper, we present a set of geometric constraints for characterizing
$N$-representability of 2-RDM. Our analysis is based on the basic
geometric property of Hilbert space, triangle inequality, and the
commutation relations of operators in fermion Fock space. These constraints
are explicitly imposed on the eigenvalues and eigenvectors of fermion
2-RDMs. Numerical examples are provided to demonstrate the evident
violation of these constraints by the variational 2-RDMs, even in
the case where the error in variational ground state energy is negligibly
small. It is also shown that, for a strongly correlated system, the
constraint violation by variational 2-RDM may contribute a large portion
of its error in ground state energy. Based on basic geometric properties
of Hilbert, our analysis is concise without direct involvement of
higher-order RDMs, and is applicable for tackling quantum marginal
problem in general.

\section*{Two-electron Reduced Density Matrix and Lie Algebra }

\subsection*{The eigenoperators of RDMs and their Lie algebra properties}

We state with some necessary notation. For a given wave function $|\Psi\rangle$
in $N$-electron Fock space, the 2-RDMs: $D$, $G$ and $Q$ matrix,
are defined as\citep{Coleman2000} \begin{subequations}

\begin{equation}
D_{ij,kl}=\langle\Psi|a_{i}^{\dagger}a_{j}^{\dagger}a_{l}a_{k}|\Psi\rangle,
\end{equation}
\begin{equation}
G_{ij,kl}=\langle\Psi|a_{i}^{\dagger}a_{j}a_{l}^{\dagger}a_{k}|\Psi\rangle,
\end{equation}
 and 
\begin{equation}
Q_{ij,kl}=\langle\Psi|a_{i}a_{j}a_{l}^{\dagger}a_{k}^{\dagger}|\Psi\rangle.
\end{equation}

\end{subequations}Here $a_{i}^{\dagger}$ and $a_{i}$ are the electron
creation and annihilation operators associated with single electron
basis 
\[
\left\{ |\phi_{i}\rangle,\,i=1,2,\cdots,L\right\} ,
\]
 respectively. The creation and annihilation operators obey the anticommutative
rules. $D$, $G$ and $Q$ are $L^{2}\times L^{2}$ matrices. They
are interconnected according to the anticommutation relations of creation
and annihilation operators. $D$ and $Q$ matrix have antisymmetric
property: $D_{ij,kl}=-D_{ji,kl}=-D_{ij,lk}$ and $Q_{ij,kl}=-Q_{ji,kl}=-Q_{ij,lk}$.

These three matrices are Hermitian and positive semidefinite, and
can be diagonalized as, \begin{subequations}
\begin{equation}
D_{ij,kl}=\sum_{n=1}^{\frac{L\left(L-1\right)}{2}}u_{ij}^{n*}\lambda_{n}^{D}u_{kl}^{n},\label{eq:eigen_D}
\end{equation}
\begin{equation}
G_{ij,kl}=\sum_{n=1}^{L^{2}}v_{ij}^{n*}\lambda_{n}^{G}v_{kl}^{n},\label{eq:eigen_G}
\end{equation}
 and 
\begin{equation}
Q_{ij,kl}=\sum_{n=1}^{\frac{L\left(L-1\right)}{2}}w_{ij}^{n*}\lambda_{n}^{Q}w_{kl}^{n}.\label{eq:eigen_Q}
\end{equation}
 \end{subequations}Here, $\lambda_{n}^{D}$, $\lambda_{n}^{G}$ and
$\lambda_{n}^{Q}$ are, respectively, the $n$th eigenvalues of $D$,
$G$ and $Q$ matrix with corresponding eigenvectors $u_{ij}^{n}$,
$v_{ij}^{n}$ and $w_{ij}^{n}$.

Using the eigenvectors, we define the eigenoperators of RDMs. For
$D$ matrix, its $n$th eigenoperator $d_{n}$ is defined by \begin{subequations}
\begin{equation}
d_{n}=\sum_{i,j=1}^{L}u_{ij}^{n}a_{j}a_{i}.\label{eq:dn}
\end{equation}
 $\mathfrak{s}_{D}$ denotes the eigenoperator set $\left\{ d_{n},\,n=1,2,\cdots,\frac{L\left(L-1\right)}{2}\right\} $.
Similarly, for $G$ and $Q$ matrices, we have 
\begin{equation}
g_{n}=\sum_{i,j=1}^{L}v_{ij}^{n}a_{j}^{\dagger}a_{i},\label{eq:gn}
\end{equation}
\[
\mathfrak{s}_{G}=\left\{ g_{n},\,n=1,2,\cdots,L^{2}\right\} ,
\]
 and

\begin{equation}
q_{n}=\sum_{i,j=1}^{L}w_{ij}^{n}a_{j}^{\dagger}a_{i}^{\dagger},\label{eq:qn}
\end{equation}
\[
\mathfrak{s}_{Q}=\left\{ q_{n},\,n=1,2,\cdots,\frac{L\left(L-1\right)}{2}\right\} .
\]

\end{subequations}Based on Eq.(\ref{eq:eigen_D}), (\ref{eq:eigen_G})
and (\ref{eq:eigen_Q}), the eigenoperators have properties: \begin{subequations}
\begin{equation}
\langle\Psi|d_{m}^{\dagger}d_{n}|\Psi\rangle=\lambda_{m}^{D}\delta_{mn},
\end{equation}
\begin{equation}
\langle\Psi|g_{m}^{\dagger}g_{n}|\Psi\rangle=\lambda_{m}^{G}\delta_{mn}
\end{equation}
 and 
\begin{equation}
\langle\Psi|q_{m}^{\dagger}q_{n}|\Psi\rangle=\lambda_{m}^{Q}\delta_{mn},
\end{equation}
\end{subequations} here $\delta_{mn}$ is kronecker delta.

The eigenoperators of RDMs are pair operators, their commutation relations
are \begin{subequations}

\begin{equation}
\left[g_{m},g_{n}\right]=\sum_{m'=1}^{L^{2}}\varGamma{}_{mn}^{m'}g_{m'},\label{eq:comm1}
\end{equation}
\begin{equation}
\left[g_{m},d_{n}\right]=\sum_{m'=1}^{\frac{L\left(L-1\right)}{2}}\Delta{}_{mn}^{m'}d_{m'},\label{eq:comm2}
\end{equation}
\begin{equation}
\left[g_{m},q_{n}\right]=\sum_{m'=1}^{\frac{L\left(L-1\right)}{2}}\Omega{}_{mn}^{m'}q_{m'},\label{eq:comm3}
\end{equation}
\begin{equation}
\left[q_{m},d_{n}\right]=\sum_{m'=1}^{L^{2}}\varTheta{}_{mn}^{m'}g_{m'},\label{eq:comm4}
\end{equation}
 and 
\begin{equation}
\left[d_{m},d_{n}\right]=\left[q_{m},q_{n}\right]=0.\label{eq:comm5}
\end{equation}

\end{subequations}

Here the coefficients are given by \begin{subequations}
\begin{equation}
\varGamma{}_{mn}^{m'}=\sum_{i,j,k=1}^{L}\left(v_{kj}^{m}v_{ik}^{n}-v_{ik}^{m}v_{kj}^{n}\right)v_{ij}^{m'*},\label{eq:comm_coef1}
\end{equation}
\begin{equation}
\Delta{}_{mn}^{m'}=-2\sum_{i,j,k=1}^{L}v_{ik}^{m}u_{kj}^{n}u_{ij}^{m'*},\label{eq:comm_coef2}
\end{equation}
\begin{equation}
\Omega{}_{mn}^{m'}=2\sum_{i,j,k=1}^{L}w_{ik}^{n}v_{kj}^{m}w_{ij}^{m'*},\label{eq:comm_coef3}
\end{equation}
 and 
\begin{equation}
\varTheta{}_{mn}^{m'}=4\sum_{i,j,k=1}^{L}\left(u_{ik}^{m}w_{kj}^{n}v_{ij}^{m'*}-\frac{1}{2N}u_{ij}^{m}w_{ji}^{n}v_{kk}^{m'*}\right).\label{eq:comm_coef4}
\end{equation}
\end{subequations} In the derivation of these commutation relations
(see Appendix for detail), we have used the facts that $u_{ij}^{m}=-u_{ji}^{m}$,
and $w_{ij}^{m}=-w_{ji}^{m}$. For Eq.(\ref{eq:comm_coef4}), we have
restricted ourselves to the $N$-electron Fock space. From these commutation
relations, we can see that the eigenoperators of RDMs form a complete
basis set of a Lie algebra. We denote this Lie algebra by $\mathfrak{h}$.
Furthermore, $\mathfrak{s}_{G}$ is a subalgebra of $\mathfrak{h}$. 

The commutators in Eq.(\ref{eq:comm1}) to (\ref{eq:comm4}) maps
the state $|\Psi\rangle$ to four unnormalized vectors in Fock space.
The length of these vectors are given by\begin{subequations}

\begin{eqnarray}
\alpha_{mn} & = & \left(\langle\Psi|\left(\left[g_{m},g_{n}\right]\right)^{\dagger}\left[g_{m},g_{n}\right]|\Psi\rangle\right)^{\frac{1}{2}}\nonumber \\
 & = & \left(\sum_{m'=1}^{L^{2}}\left|\Gamma_{mn}^{m'}\right|^{2}\lambda_{m'}^{G}\right)^{\frac{1}{2}},\label{eq:def_alpha}
\end{eqnarray}

\begin{eqnarray}
\beta_{mn} & = & \left(\langle\Psi|\left(\left[q_{m}d_{n}\right]\right)^{\dagger}\left[q_{m}d_{n}\right]|\Psi\rangle\right)^{\frac{1}{2}}\nonumber \\
 & = & \left(\sum_{m'=1}^{L^{2}}\left|\varTheta{}_{mn}^{m'}\right|^{2}\lambda_{m'}^{G}\right)^{\frac{1}{2}},\label{eq:def_beta}
\end{eqnarray}

\begin{align}
\gamma_{mn} & =\left(\langle\Psi|\left(\left[g_{m}d_{n}\right]\right)^{\dagger}\left[g_{m}d_{n}\right]|\Psi\rangle\right)^{\frac{1}{2}}\nonumber \\
 & =\left(\sum_{m'=1}^{\frac{L\left(L-1\right)}{2}}\left|\Delta{}_{mn}^{m'}\right|^{2}\lambda_{m'}^{D}\right)^{\frac{1}{2}},\label{eq:def_gamma}
\end{align}
 and
\begin{align}
\zeta_{mn} & =\left(\langle\Psi|\left(\left[g_{m}q_{n}\right]\right)^{\dagger}\left[g_{m}q_{n}\right]|\Psi\rangle\right)^{\frac{1}{2}}\nonumber \\
 & =\left(\sum_{m'=1}^{\frac{L\left(L-1\right)}{2}}\left|\Omega{}_{mn}^{m'}\right|^{2}\lambda_{m'}^{Q}\right)^{\frac{1}{2}}.\label{eq:def_zeta}
\end{align}

\end{subequations}These lengths will be used later for verifying
the effectiveness of N-representability constraints. 

\subsection*{The null eigenoperators of RDMs }

An operator is called null operator of a wave function if it maps
the wave function to null vector. For a given wave function, the commutator
of two null operators must be a null operator. Therefore, all the
null operators of a given wave function form a Lie algebra. 

In the Lie algebra $\mathfrak{h}$, we define a subset $\mathfrak{n=}\left\{ p\in\mathfrak{h}:\forall\,p|\Psi\rangle=0\right\} $.
Apparently, $\mathfrak{n}$ is a subalgera of $\mathfrak{h}$. For
any operator 
\begin{equation}
p=\sum_{i,j=1}^{L}\left(r_{ij}a_{j}a_{i}+s_{ij}a_{j}^{\dagger}a_{i}+t_{ij}a_{j}^{\dagger}a_{i}^{\dagger}\right)\label{eq:pair op}
\end{equation}
 in $\mathfrak{n}$, we have 
\begin{eqnarray*}
0 & = & \langle\Psi|p^{\dagger}p|\Psi\rangle\\
 & = & \sum_{ij,kl}\left(r_{ij}^{*}r_{kl}\langle\Psi|a_{i}^{\dagger}a_{j}^{\dagger}a_{l}a_{k}|\Psi\rangle+s_{ij}^{*}s_{kl}\langle\Psi|a_{i}^{\dagger}a_{j}a_{l}^{\dagger}a_{k}|\Psi\rangle+t_{ij}^{*}t_{kl}\langle\Psi|a_{i}a_{j}a_{l}^{\dagger}a_{k}^{\dagger}|\Psi\rangle\right)\\
 & = & \sum_{ij,kl}\left(r_{ij}^{*}r_{kl}D_{ij,kl}+s_{ij}^{*}s_{kl}G_{ij,kl}+t_{ij}^{*}t_{kl}Q_{ij,kl}\right)
\end{eqnarray*}
 which implies 
\begin{equation}
\sum_{ij,kl}r_{ij}^{*}D_{ij,kl}r_{kl}=\sum_{ij,kl}s_{ij}^{*}G_{ij,kl}s_{kl}=\sum_{ij,kl}t_{ij}^{*}Q_{ij,kl}t_{kl}=0,\label{eq:null space}
\end{equation}
since $D$, $G$ and $Q$ matrix are positive semidefinite. Eq.(\ref{eq:null space})
shows that the vector $r_{ij}$, $s_{ij}$ and $t_{ij}$ are in the
null space of $D$, $G$ and $Q$ matrix, respectively. They can be
expanded by linear combinations of the eigenvectors in the null space
of RDMs as\begin{subequations}
\begin{equation}
r_{ij}=\sum_{m=1}^{N_{null}^{D}}c_{m}^{r}u_{ij}^{m},\label{eq:exp_r}
\end{equation}
\begin{equation}
s_{ij}=\sum_{m=1}^{N_{null}^{G}}c_{m}^{s}v_{ij}^{m},\label{eq:exp_s}
\end{equation}
 and 
\begin{equation}
t_{ij}=\sum_{m=1}^{N_{null}^{Q}}c_{m}^{t}w_{ij}^{m},\label{eq:exp_t}
\end{equation}
\end{subequations}here, the expansion coefficients are given by $c_{m}^{r}=\sum_{i,j=1}^{L}u_{ij}^{m*}r_{ij}$,
$c_{m}^{s}=\sum_{i,j=1}^{L}v_{ij}^{m*}s_{ij}$ and $c_{m}^{t}=\sum_{i,j=1}^{L}w_{ij}^{m*}t_{ij}$.
$N_{null}^{D}$, $N_{null}^{G}$ and $N_{null}^{Q}$ are the dimension
of the null spaces of $D$, $G$ and $Q$ matrix, respectively. $m$
is the index for the corresponding null eigenvector of RDMs.

Substituting Eq.(\ref{eq:exp_r}), (\ref{eq:exp_s}) and (\ref{eq:exp_t})
into Eq.(\ref{eq:pair op}), we have the expansion for operator $p$
as
\begin{equation}
p=\sum_{m=1}^{N_{null}^{D}}c_{m}^{r}d_{m}+\sum_{m=1}^{N_{null}^{G}}c_{m}^{s}g_{m}+\sum_{m=1}^{N_{null}^{Q}}c_{m}^{t}q_{m}\label{eq:exp_null}
\end{equation}

We call an eigenoperator the null eigenoperator of RDM if its associated
eigenvector is in the null space of RDM. Eq.(\ref{eq:exp_null}) indicates
that the set of all null eigenoperators must from a complete basis
set of the subalgebra $\mathfrak{n}$. Similarly, it can be shown
that the operator vector space $\mathfrak{n}_{G}$ spanned by the
null eigenoperators of $G$ matrix must be a subalgebra of $\mathfrak{n}$.

\section*{The Constraints on the Null Spaces of 2-RDMs}

The requirement that all null eigenoperators form a Lie algebra impose
a set of nontrivial constraints on the null spaces of 2-RDMs. For
two null eigenoperators of the $G$ matrix: $g_{m}$, $g_{n}$ and
their commutator $\left[g_{m},g_{n}\right]$, using Eq.(\ref{eq:gn}),
we have (see Appendix for detail)
\begin{equation}
\left[g_{m},g_{n}\right]|\Psi\rangle=\sum_{i,j=1}^{L}v_{ij}^{mn}a_{j}^{\dagger}a_{i}|\Psi\rangle,
\end{equation}
 here 
\begin{equation}
v_{ij}^{mn}=\sum_{k=1}^{L}\left(v_{kj}^{m}v_{ik}^{n}-v_{ik}^{m}v_{kj}^{n}\right).
\end{equation}

The length of vector $\left[g_{m},g_{n}\right]|\Psi\rangle$ vanishes,
and we have
\begin{align}
0= & \langle\Psi|\left(\left[g_{m},g_{n}\right]\right)^{\dagger}\left[g_{m},g_{n}\right]|\Psi\rangle\nonumber \\
= & \sum_{i,j,k,l=1}^{L}\left(v_{ij}^{mn}\right)^{*}G_{ij,kl}v_{kl}^{mn},\label{eq:G_null_space}
\end{align}
 which indicates that $\boldsymbol{v}^{mn}$ must be in the null space
of the $G$ matrix. This requirement imposes a constraint on the null
space of the $G$ matrix.

If $d_{m}$ and $q_{n}$ are, respectively, the null eigenoperators
of $D$ and $Q$ matrices, we can derive the constraint on the null
spaces of $D$ and $Q$ matrices, 
\begin{equation}
0=\sum_{i,j,k,l=1}^{L}\left(\overline{v}_{ij}^{mn}\right)^{*}G_{ij,kl}\overline{v}_{kl}^{mn},
\end{equation}
here 
\begin{equation}
\overline{v}_{ij}^{mn}=4\left(\sum_{k=1}^{L}u_{ik}^{m}w_{kj}^{n}-\frac{1}{2N}\sum_{k,k'=1}^{L}u_{kk'}^{m}w_{k'k}^{n}\delta_{ij}\right).
\end{equation}

More constraints on the null spaces can be found using the commutation
relations, Eq.(\ref{eq:comm2}) and (\ref{eq:comm3}). They are 
\begin{equation}
0=\sum_{i,j,k,l=1}^{L}\left(\overline{u}{}_{ij}^{mn}\right)^{*}D_{ij,kl}\overline{u}{}_{kl}^{mn}
\end{equation}
 with 
\begin{equation}
\overline{u}{}_{ij}^{mn}=-2\sum_{k=1}^{L}v_{ik}^{m}u_{kj}^{n},
\end{equation}
 and 
\begin{equation}
0=\sum_{i,j,k,l=1}^{L}\left(\overline{w}{}_{ij}^{mn}\right)^{*}Q_{ij,kl}\overline{w}{}_{kl}^{mn}
\end{equation}
 with
\begin{equation}
\overline{w}{}_{ij}^{mn}=2\sum_{k=1}^{L}w_{ik}^{n}v_{kj}^{m}.
\end{equation}

These constraints are equivalent to the conditions that the lengths
given in Eq.(\ref{eq:def_alpha}) to (\ref{eq:def_zeta}) vanish for
the commutators of two null eigenoperators. They show why the positive
semidefinite conditions on 2-RDMs are not strong enough to ensure
$N$-rerepresentability.

\section*{Numerical Verification of Constraint Effectiveness}

To examine the effectiveness of the constraints on the null spaces
numerically, we employ variational 2-RDM method to obtain approximated
2-RDMs of ground state. In variational 2-RDM method, the total energy
of a system is a function of $D$ matrix, $E=tr\left(KD\right)$.
Here, $K$ is the Hamiltion matrix of the system. The ground state
energy of the system is obtained by variationally minimizing the total
energy with respect to the $D$ matrix under the restriction of $N$-representability
constraints. The currently available constraints are not restrictive
enough to ensure the $N$-representability of the variational 2-RDM
($D_{var}$). Therefore, the variational energy ($E_{var})$ provides
a low-boundary estimation of ground state energy. Variational 2-RDM
method has been utilized routinely in the past to demonstrate the
effectiveness of $N$-representability conditions.\citep{Nakata2001,Zhao2004,Neck2007,Nakata2008,Shenvi2010,Johnson2013,Mazziotti2016}
In numerical tests, we perform variational 2-RDM method first to obtain
$E_{var}$ and $D_{var}$, and calculate the variational $G$ and
$Q$ matrices ($G_{var}$ and $Q_{var}$) from $D_{var}$. Then, the
eigenvalues and eigenvectors of $D_{var}$, $G_{var}$ and $Q_{var}$
are used to check whether the constraints given in previous section
are held by the variational 2-RDMs. In the variational 2-RDM calculations,
we have applied $D$, $Q$, $G$, $T1$, $T2$ and $T2'$ conditions,\citep{Zhao2004,Nakata2008}
which, to the best of our knowledge, are the most restrictive constraints
currently available for practical computation.

In this section, variational RDMs are calculated for several systems:
one-dimensional Hubbard model, diatomic molecule LiH and two random-matrix
Hamiltonians with free spin. The numerical results are summarized
in Table 1. In order to reduce the number of variational variables
and to improve numerical accuracy, the linear equalities derived from
the symmetries of specific systems are solved explicitly before variational
calculation. For comparison, the exact RDMs are also calculated by
the full configuration interaction method (FCI). Variational calculations
are carried out using a SDP software, Sedumi 1.3.\citep{Sturm1999}
According to the ``prec'' parameter in Sedumi output, the numerical
accuracy is about $10^{-9}$ in variational calculations, so we regard
any value in $(-10^{-9},10^{-9})$ as numerical zero. 

\subsection*{Hubbard Model}

Hubbard model is a prototype system for studying strong correlated
electrons.\citep{Dagotto1994} Here, the system is a 6-site half-filled
1D Hubbard model with periodic boundaries, $t=1$ and varying $U$.
For $U/t=10$, the ground state energies obtained by the FCI and variational
2-RDM method are $E_{exact}=-1.664362733287$ and $E_{var}=-1.695384327725$,
respectively. Compared to $E_{exact}$, the energy deviation $\Delta E=-0.031021594438$.
These energies are consistent with the previous studies on this model.\citep{Nakata2008} 

\setlength{\extrarowheight}{4pt}

\begin{table}[h]
\caption{Summary of numerical tests for four systems: Hubbard Model with $U/t=10$,
diatomic molecule LiH, and two random matrix Hamiltonians. $L$ and
$N_{e}$ are the number of single particle basis and electrons in
system, respectively. The number of null eigenoperators (NEO) for
variational $D_{var}$, $G_{var}$ and $Q_{var}$ matrices are shown
with the number for exact RDMs given in parentheses for comparison.
$\Delta E$ is the deviation of variational ground state energy from
the exact value. $\Delta E_{null}/\left|\Delta E\right|$ is an estimation
for the contribution of the null space of $D_{var}$ to the ground-state-energy
deviation $\Delta E$. $I_{\alpha}$, $I_{\beta}$, $I_{\gamma}$
and $I_{\zeta}$ are four descriptors to quantify the extent that
the $N$-representability constraints are violated by variational
RDMs. For $N$-representable RDMs, these descriptors should vanish.}

\begin{tabular}{|c|c|c|c|c|}
\hline 
 & Hubbard Model & LiH & Random 1 & Random 2\tabularnewline
\hline 
\hline 
$L$ & 12 & 12 & 12 & 12\tabularnewline
\hline 
$N_{e}$ & 6 & 4 & 6 & 6\tabularnewline
\hline 
No. of NEOs of $D$ & 6 (1) & 2 (0) & 1 (0) & 4 (0)\tabularnewline
\hline 
No. of NEOs of $G$ & 8 (3) & 5 (13) & 7 (3) & 7 (3)\tabularnewline
\hline 
No. of NEOs of $Q$ & 6 (1) & 1 (0) & 3 (0) & 1 (0)\tabularnewline
\hline 
$\Delta E$ & $-3.1\times10^{-2}$ & $-1.7\times10^{-8}\,E_{h}$ & $-1.3\times10^{-1}$ & $-1.3\times10^{-2}$\tabularnewline
\hline 
$\Delta E_{null}/\left|\Delta E\right|$ & $42\%$ & $\ensuremath{-67\%}$$^{a}$ & $4.2\%$ & $6.5\%$\tabularnewline
\hline 
$I_{\alpha}$ & $0.78\times10^{-2}$ & 0 & $0.19\times10^{-1}$ & $0.78\times10^{-2}$\tabularnewline
\hline 
$I_{\beta}$ & $0.21\times10^{-1}$ & $0.30\times10^{-3}$ & $0.43\times10^{-1}$ & $0.21\times10^{-1}$\tabularnewline
\hline 
$I_{\gamma}$ & $0.15\times10^{-1}$ & \textbf{$0.87\times10^{-4}$} & $0.16\times10^{-1}$ & $0.15\times10^{-1}$\tabularnewline
\hline 
$I_{\zeta}$ & $0.14\times10^{-1}$ & $0.12\times10^{-3}$ & $0.46\times10^{-1}$ & $0.14\times10^{-1}$\tabularnewline
\hline 
\end{tabular}

$a$ this value is problematic because, for LiH, the $\Delta E_{null}$
and $\left|\Delta E\right|$ are both on the order of $10^{-8}$,
and are close to the numerical accuracy of variational calculation
(\textasciitilde$10^{-9}$).
\end{table}

For the exact RDMs in Table A1 (see Appendix), the eigenvalues of
$D_{exact}$ and $Q_{exact}$ matrix have one null eigenoperator each,
corresponding to the pseudospin operator and its conjugate transpose,
which is known for a half-filled Hubbard model.\citep{Zhang1990}
The three null eigenvalues of the $G_{exact}$ matrix are corresponding
to three spin operators, $s_{x}$, $s_{y}$ and $s_{z}$ because the
ground state of the half-filled Hubbard model is a singlet. 

Comparing with $D_{exact}$, there are five more null eigenoperators
for $D_{var}$ (Table 1). The ground state energy $E=tr\left(KD\right)$
with $K$ being the Hamiltonian matrix, so the 6-dimensional null
space of $D_{var}$ has no contribution to the variational ground
state energy, $E_{var}$. To roughly assess how much the null space
of $D_{var}$ contributes to the deviation of ground state energy
$\Delta E$, we first project $D_{exact}$, onto the 6D null space,
and then calculate the energy contribution of the projected $D$ matrix.
Let $P=\sum_{n=1}^{6}\mathbf{u}_{n}\mathbf{u}_{n}^{T}$, be the project
matrix, here $\mathbf{u}_{n},\,n=1,2,\cdots,6$ are the eigenvectors
in the null space of $D_{var}$. Then, the energy contribution 
\begin{eqnarray}
\Delta E_{null} & = & tr(KPD_{exact}P)\\
 & = & 0.013134139307\nonumber 
\end{eqnarray}
The positive value of $\Delta E_{null}$ indicates that the erroneous
null space causes the underestimation of ground state energy by $D_{var}$.
Its contribution to the energy deviation is quite large, $\Delta E_{null}/\left|\Delta E\right|\approx42\%$,
even though its dimension is small. Fig. 1(a) shows that the null
space of $D_{var}$ has overlap with not only the low-lying eigenvectors
of $D_{exact}$ but also the high-lying ones. This may explain its
large contribution to the energy deviation. Furthermore, the contribution
of the erroneous null space is correlated with the deviation of variational
energy for the Hubbard model with varying $U/t$ (Fig.1(b)). 

\begin{figure}[h]

\appendix
\includegraphics[scale=0.1]{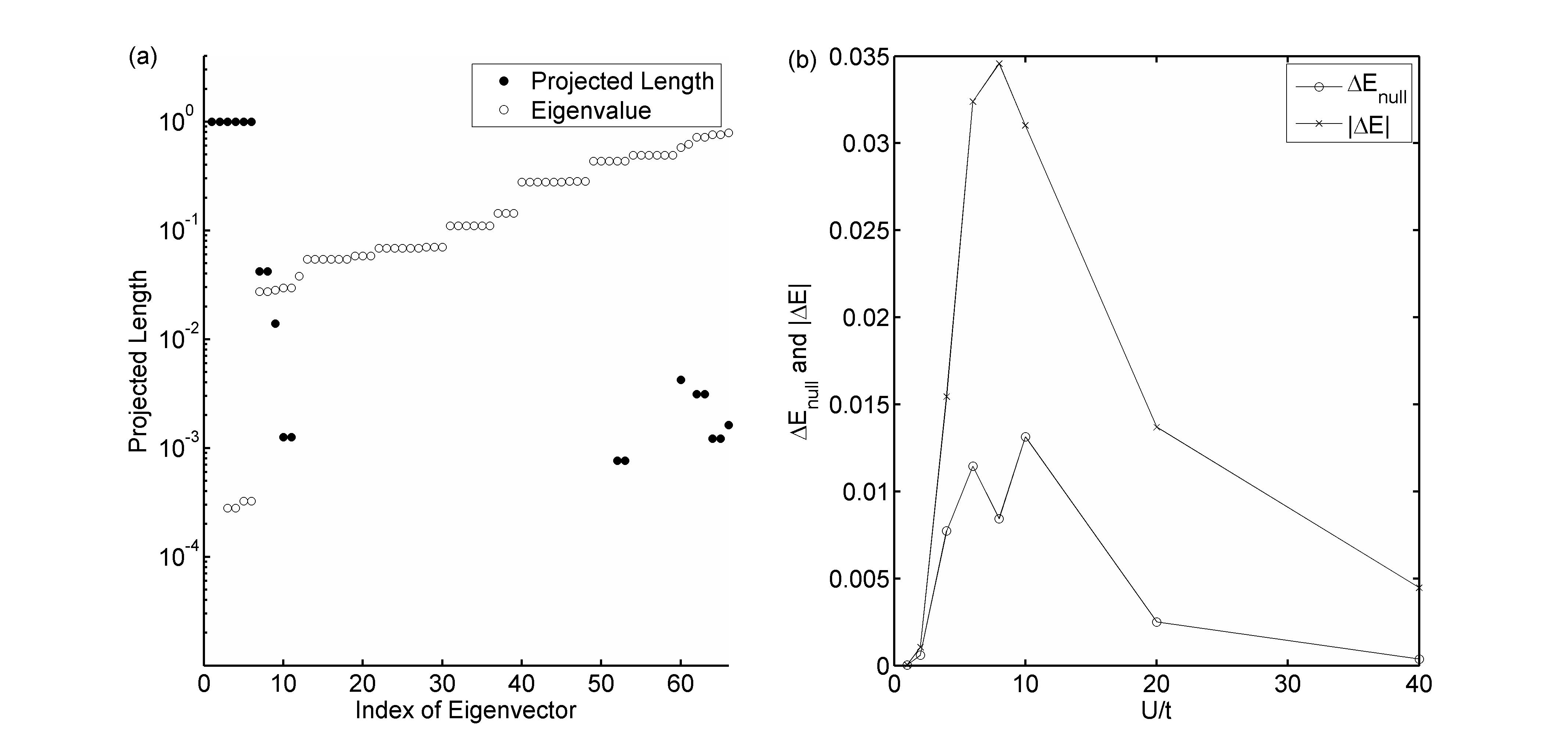}

\caption{(a) The projection length (filled circle) of the eigenvector of $D_{exact}$
in the null space of $D_{var}$ together with the corresponding eigenvalues
(open circle). The eigenvalues of $D_{exact}$ are sorted in ascending
order. The results are for 1D Hubbard model with $U/t=10$. The null
space of $D_{var}$ has overlap with 10 low-lying and 8 high-lying
eigenvectors of $D_{exact}$. (b) The contribution of the $D_{var}$
null space to the deviation of variational total energy for 1D Hubbard
model with $U/t=1,2,4,6,8,10,20,40$, respectively. The contribution
is largely correlated with the deviation of variational total energy.}
\end{figure}

$G_{var}$ has five more null eigenvalues than $G_{exact}$. As discussed
in previous section, if $G_{var}$ is $N$-representable, the eight
corresponding null eigenoperators must form the complete basis set
of a Lie algebra. That is, For two null eigenoperators $g_{m}$ and
$g_{n}$ in $\mathfrak{n}_{G}$, $\left[g_{m},g_{n}\right]|\Psi\rangle=0$. 

\begin{figure}[h]
\includegraphics[clip,scale=0.15]{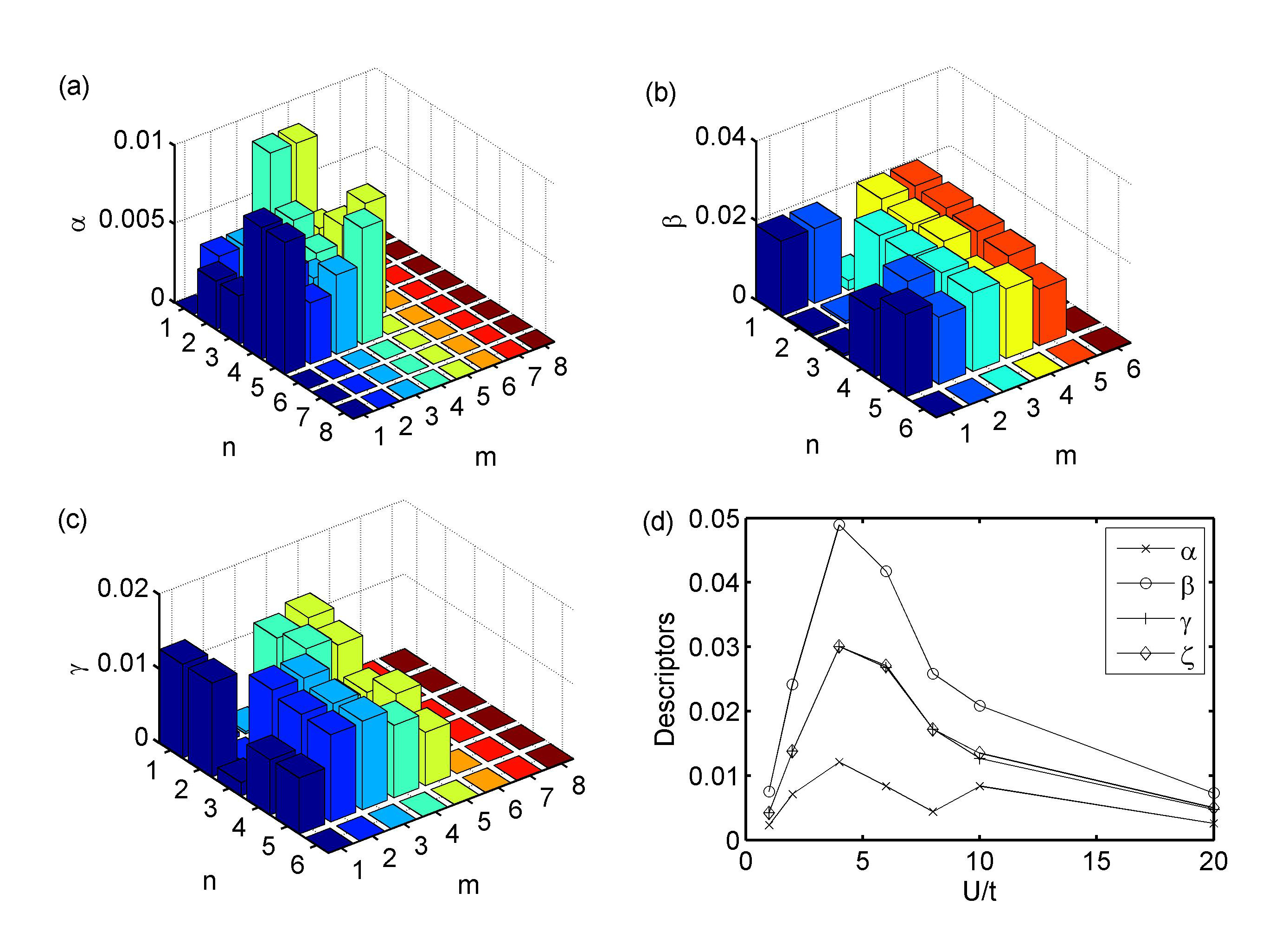}

\caption{The constraint test on the variational RDMs of Hubbard model. (a)
The length of vector $\left[g_{m},g_{n}\right]|\Psi\rangle$ with
$m,n=1,2,\cdots,8$ corresponding to the eight null eigenoperators
of $G_{var}$. (b) The length of vector $\left[d_{m},q_{n}\right]|\Psi\rangle$
with $m,n=1,2,\cdots,6$ corresponding to the six null eigenoperators
of $D_{var}$ and $Q_{var}$. (c) The length of vector $\left[g_{m},d_{n}\right]|\Psi\rangle$
with $m=1,2,\cdots,8$ and $n=1,2,\cdots,6$ corresponding to the
eight null eigenoperators of $G_{var}$ and six for $Q_{var}$. The
maximal lengths in (a), (b) and (c) are, respectively, about 0.0078,
0.021 and 0.015. Apparently, many of the commutators are not null
operators of the state $|\Psi\rangle$, therefore $D_{var}$, $G_{var}$
and $Q_{var}$ are not $N$-representable. (d) The constraint violation
descriptors $I_{\alpha}$, $I_{\beta}$, $I_{\gamma}$ and $I_{\zeta}$
for the Hubbard model with $U/t=1,2,4,6,8,10,20,40$. The maximum
of four descripters is around $U/t=4$. While, the maximal deviation
of variational energy is at $U/t=8$ (see Fig. 1(b)). Therefore, the
structure deviation of the variational 2-RDM is not correlated with
the error of the total energy.}
\end{figure}

We use Eq.(\ref{eq:def_alpha}) and (\ref{eq:def_beta}) to verify
whether the variational 2-RDMs are $N$-representable. Fig. 2(a) shows
$\alpha_{mn}$ for the null eigenoperators of $G_{var}$. There are
multiple non-vanishing values. Therefore, the null eigenoperators
of $G_{var}$ do not form a closed subalgebra, and $G_{var}$ is not
$N$-representable.

$Q_{var}$ has five more null eigenvalues than $Q_{exact}$. If $Q_{var}$
and $D_{var}$ are $N$-representable, the commutator of their null
eigenoperators must be a null eigenoperator in $\mathfrak{n}_{G}$.
Fig.2(b) shows the the $\beta_{mn}$ evaluated from the null eigenoperators
of $Q_{var}$ and $D_{var}$. The values are in the order of $10^{-2}$
and, clearly, violate the $N$-representability constraint. The constraint
violation is aslo prominent for the commutors of the null eigenoperators
for $G_{var}$ and $D_{var}$ (Fig. 2(c)).

The numerical results for Hubbard model show that the constraints
impose strong restrictions on the null spaces of 2-RDMs. To quantify
the degree of constraint violation by variational 2-RDMs, we introduce
a descriptor $I_{\alpha}$ defined as the maximum value of $\alpha_{mn}$
from the null eigenoperators of $G$. A large value of descriptor
suggests strong violation. Similarly, we may define descriptors $I_{\beta}$,
$I_{\gamma}$ and $I_{\zeta}$ for the maximum value of $\beta_{mn}$,
$\gamma_{mn}$ and $\zeta_{mn}$, respectively. The four descriptors
of the Hubbard model with $U/t=10$ are shown in Table 1.

$N$-representability requires the constraints on the null spaces
of 2-RDMs:
\begin{equation}
I_{\alpha}=I_{\beta}=I_{\gamma}=I_{\zeta}=0.\label{eq:eq_const}
\end{equation}
 Fig. 2(d) shows the trend of constraint violation by variational
2-RDMs as $U/t$ of Hubbard model varying. In general, we can see
to $I_{\beta}>I_{\gamma}\approx I_{\zeta}>I_{\alpha}$. The maximum
of these descriptors is around $U/t=4$, which is different from that
for the variational energy deviation (around $U/t=8$ as shown in
Fig. 1(b)). 

\subsection*{LiH and random-matrix Hamiltonians}

The violation of the null space constraint seems general for variational
2-RDM. For the diatomic molecule LiH in its equilibrium configuration,
the variational method can provide the very accurate estimation of
ground state energy with $\Delta E\sim2.0\times10^{-8}$ (see Table
A2). However, the dimensions of the null spaces of variational 2-RDMs
are very different from that of the exact 2-RDMs (see Table A2 and
Table 1), which indicates the disparity in the Lie algebra structure
of their null eigenoperators. The descriptor $I_{\beta}=0.30\times10^{-3}$,
which is about 5 order of magnitude larger than the numerical accuracy
of our variational calculation.

In order to have a rough idea how often the variational 2-RDM method
may predict erroneous null spaces of RDMs, we have applied the method
to five Hamiltonians with randomly generated numbers in spatial degree
of freedom. The erroneous null spaces have been found in all five
cases. The results for two of them are shown in Table A3 and A4. As
summarized in Table 1, the exact RDMs of the ground states have 3
null eigenoperators corresponding to 3 spin operators. While, the
variational RDMs have more null eigenoperators. The four descriptors
are not vanishing for these RDMs. 

From Table 1, we can have two interesting observations. In contrast
to Hubbard model, a strongly correlated system, the $\Delta E_{null}/\left|\Delta E\right|$
(\textasciitilde 5\%) of the random systems is quite small, and the
null space of $D_{var}$ has very small contribution to the ground
state energy deviation $\Delta E$. We belive, $\Delta E_{null}/\left|\Delta E\right|$
(-67\%) for LiH is problematic because both $\left|\Delta E_{null}\right|$
and $\left|\Delta E\right|$ are on the of $10^{-8}$ and close to
numerical accuracy $1.0\times10^{-9}$. For variation RDMs, the descriptor
$I_{\beta}$ , $I_{\gamma}$ and $I_{\zeta}$ are usually larger than
$I_{\alpha}$. This may suggest that the constraint on the null spaces
of $D$ and $Q$ matrices is stronger than that on $G$ matrix. Apparently,
more numerical studies are needed in future in order to tell whether
the observations are general.

\section*{Inequality Constraints on the Whole Eigenspaces of RDMs}

The constraints presented so far are only applied to the null spaces
of RDMs. To derive the constraints covering the whole eigenspaces
of RDMs, we first use commutation relation of two operators to have
a vector equation: 
\[
p_{1}p_{2}|\Psi\rangle-p_{2}p_{1}|\Psi\rangle=p_{3}|\Psi\rangle,
\]
 here $p_{1}$, $p_{2}$ and $p_{3}$ are three pair operators in
$\mathfrak{h}$ satisfying $p_{3}=\left[p_{1},p_{2}\right]$. Using
triangle inequality, we have 
\begin{equation}
\left|p_{3}|\Psi\rangle\right|\leq\left|p_{1}p_{2}|\Psi\rangle\right|+\left|p_{2}p_{1}|\Psi\rangle\right|\label{eq:triangle ineq}
\end{equation}
 with $\left|p_{1}p_{2}|\Psi\rangle\right|^{2}=\langle\Psi|p_{2}^{\dagger}p_{1}^{\dagger}p_{1}p_{2}|\Psi\rangle$,
$\left|p_{2}p_{1}|\Psi\rangle\right|^{2}=\langle\Psi|p_{1}^{\dagger}p_{2}^{\dagger}p_{2}p_{1}|\Psi\rangle$
and $\left|p_{3}|\Psi\rangle\right|^{2}=\langle\Psi|p_{3}^{\dagger}p_{3}|\Psi\rangle$.
To find the upper bound of $\langle\Psi|p_{2}^{\dagger}p_{1}^{\dagger}p_{1}p_{2}|\Psi\rangle$,
let $|\phi\rangle=\frac{p_{2}|\Psi\rangle}{\left|p_{2}|\Psi\rangle\right|}$
and insert it into the inner product. W have
\begin{align}
\langle\Psi|p_{2}^{\dagger}p_{1}^{\dagger}p_{1}p_{2}|\Psi\rangle & =\langle\Psi|p_{2}^{\dagger}|\phi\rangle\langle\phi|p_{1}^{\dagger}p_{1}|\phi\rangle\langle\phi|p_{2}|\Psi\rangle\nonumber \\
 & =\langle\phi|p_{1}^{\dagger}p_{1}|\phi\rangle\langle\Psi|p_{2}^{\dagger}|\phi\rangle\langle\phi|p_{2}|\Psi\rangle\nonumber \\
 & =\langle\phi|p_{1}^{\dagger}p_{1}|\phi\rangle\langle\Psi|p_{2}^{\dagger}p_{2}|\Psi\rangle\nonumber \\
 & \leq c_{1}\langle\Psi|p_{2}^{\dagger}p_{2}|\Psi\rangle\label{eq:ineq1}
\end{align}
 here $\langle\phi|p_{1}^{\dagger}p_{1}|\phi\rangle$ is upper bounded
by $c_{1}$. Similarly, we can have 
\begin{equation}
\langle\Psi|p_{1}^{\dagger}p_{2}^{\dagger}p_{2}p_{1}|\Psi\rangle\leq c_{2}\langle\Psi|p_{1}^{\dagger}p_{1}|\Psi\rangle\label{eq:ineq2}
\end{equation}
 with $|\varphi\rangle=\frac{p_{1}|\Psi\rangle}{\left|p_{1}|\Psi\rangle\right|}$
and $\langle\varphi|p_{2}^{\dagger}p_{2}|\varphi\rangle$ is upper
bounded by $c_{2}$. From Eq.(\ref{eq:triangle ineq}), (\ref{eq:ineq1})
and (\ref{eq:ineq2}), we have

\begin{equation}
\left|p_{3}|\Psi\rangle\right|\leq c_{1}^{\frac{1}{2}}\left|p_{2}|\Psi\rangle\right|+c_{2}^{\frac{1}{2}}\left|p_{1}|\Psi\rangle\right|\label{eq:ineq_const}
\end{equation}
Substituting $p_{1}$ and $p_{2}$ in Eq.(\ref{eq:ineq_const}) by
two eigenoperators, and using the commutation relations: Eq.(\ref{eq:comm1})
to (\ref{eq:comm_coef4}), the constraints on the eigenspaces of 2-RMDs
are given by\begin{subequations}
\begin{equation}
\alpha_{mn}\leq\left(\overline{\lambda}_{N}^{G}\right)^{\frac{1}{2}}\left[\left(\lambda_{m}^{G}\right)^{\frac{1}{2}}+\left(\lambda_{n}^{G}\right)^{\frac{1}{2}}\right],\label{eq:ineq_alpha}
\end{equation}

\begin{equation}
\beta_{mn}\leq\left(\overline{\lambda}_{N-2}^{Q}\right)^{\frac{1}{2}}\left(\lambda_{m}^{D}\right)^{\frac{1}{2}}+\left(\overline{\lambda}_{N+2}^{D}\right)^{\frac{1}{2}}\left(\lambda_{n}^{Q}\right)^{\frac{1}{2}},\label{eq:ineq_beta}
\end{equation}

\begin{equation}
\gamma_{mn}\leq\left(\overline{\lambda}_{N}^{D}\right)^{\frac{1}{2}}\left(\lambda_{m}^{G}\right)^{\frac{1}{2}}+\left(\overline{\lambda}_{N-2}^{G}\right)^{\frac{1}{2}}\left(\lambda_{n}^{D}\right)^{\frac{1}{2}},\label{eq:ineq_gamma}
\end{equation}
 and 
\begin{equation}
\zeta_{mn}\leq\left(\overline{\lambda}_{N}^{Q}\right)^{\frac{1}{2}}\left(\lambda_{m}^{G}\right)^{\frac{1}{2}}+\left(\overline{\lambda}_{N+2}^{G}\right)^{\frac{1}{2}}\left(\lambda_{n}^{Q}\right)^{\frac{1}{2}}.\label{eq:ineq_zeta}
\end{equation}
\end{subequations}Here $\overline{\lambda}_{N}^{D}$, $\overline{\lambda}_{N}^{G}$
and $\overline{\lambda}_{N}^{Q}$ are, respectively, the upper bounds
for the eigenvalues of $G,$$D$ and $Q$ matrices for a $N$-electron
state. We refer the four constraints as $\alpha,$ $\beta$, $\gamma$
and $\zeta$ conditions. They are necessary $N$-representability
conditions. The constraints on the null spaces, Eq.(\ref{eq:eq_const}),
are special cases of above inequalities where the eigenvalues on the
right side vanish. From Eq.(\ref{eq:ineq_const}), we can see that
these constraints are of geometric nature. 

\begin{figure}[h]
\includegraphics[clip,scale=0.12]{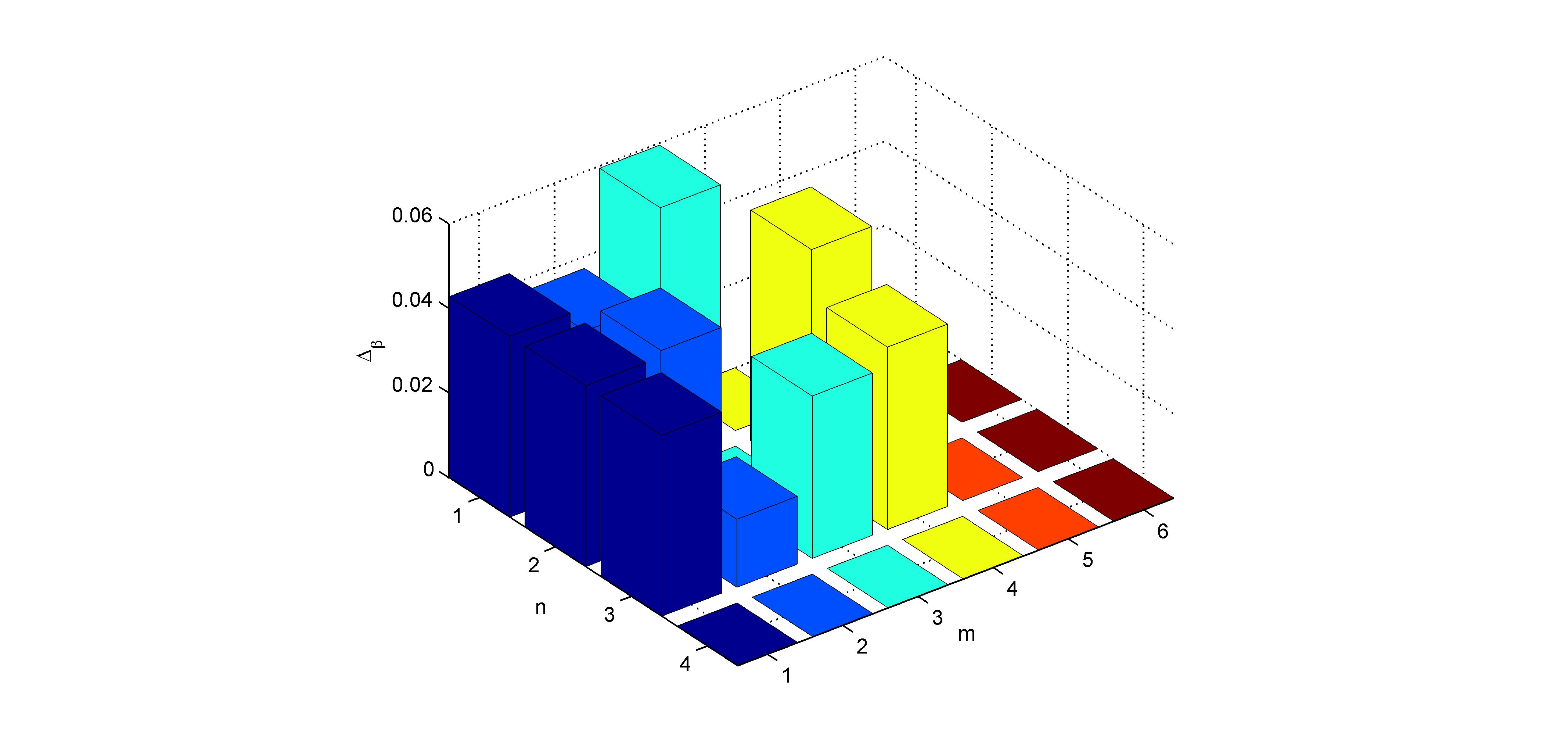}

\caption{An example for violation of the inquality constraints. The system
is a random-matrix Hamiltonians (Table A3). $\Delta_{\beta}$ is an
descriptor of constraint violation (see the definition in text). $\Delta_{\beta}>0$
indicates constraint violation. $m$ and $n$ are the indices of the
eigenvalues of $D$ and $Q$, respectively. The three dark blue bars
indicate the violation of the equality constraint Eq.(\ref{eq:eq_const})
by the variational 2-RDMs. The other bars show the explicit violation
of inequality constraints. }
\end{figure}

Fig. 3 show an example where a variationl 2-RDM violates the inequality
constants Eq.(\ref{eq:ineq_beta}). The system is a random-matrix
Hamiltonians (Table A3). The descriptor of constraint violation is
$\Delta_{\beta}\left(m,n\right)=\beta_{mn}-\left[\frac{L-N+2}{2}\right]^{\frac{1}{2}}\left(\lambda_{m}^{D}\right)^{\frac{1}{2}}-\left[\frac{N+2}{2}\right]^{\frac{1}{2}}\left(\lambda_{n}^{Q}\right)^{\frac{1}{2}}.$
$\Delta_{\beta}>0$ indicates constraint violation. we have set $\overline{\lambda}_{N+2}^{D}=\left[\frac{N+2}{2}\right]$
and $\overline{\lambda}_{N-2}^{Q}=\left[\frac{L-N+2}{2}\right]$ ,
respectively, the universal upper bounds for the eigenvalues of $D$
and $Q$ matrices.\citep{Sasaki1965} $m$ and $n$ are the indices
of the eigenvalues of $D$ and $Q$, respectively. As shown in Table
A3, $D_{var}$ has one null eigenvalue corresponding to $m=1$, and
$Q_{var}$ has three corresponding to $n=1,2,3$. Therefore, the three
dark blue bars in Fig.3 indicate the violation of the equality constraint
Eq.(\ref{eq:eq_const}). The other bars show the explicit violation
of inequality constraints. 

\section*{Discussion and Conclusion}

Even though the derivation of geometric constraints is started with
a wave function in $N$-electron Fock space, these constraints are
actually ensemble $N$-representability conditions because any mixed
state can be mapped onto a pure state in a larger space (known as
purification of mixed state in quantum information).\citep{Kleinmann2006} 

The null eigenoperators of 2-RDMs carry the information about the
conserved observables of the underlying many-electron state. Eq.(\ref{eq:eq_const})
imposes restrictions on the null eigenspace of 2-RDMs to ensure the
commmutative compatibility of these observables. More ganerally, if
$|\Psi\rangle$ is an eigenstate of a two-electron operator, $H=\sum_{ij,kl}K_{ij,kl}a_{i}^{\dagger}a_{j}^{\dagger}a_{l}a_{k}$,
then we can generate a set of two-electron null operators by $\left\{ \left[\mathfrak{n}_{G},H\right],\left[\mathfrak{n}_{G},\left[\mathfrak{n}_{G},H\right]\right],\cdots\right\} $,
which provides additional constraints on the 2-RDMs. Prediction of
fractional charges and fractiona spins is an indication of insufficient
constraint on the conserved observables in various RDM-based electronic
structure methods.

The results shown in Fig. 2(d) and for LiH suggests that, for variational
2-RDM method, a smaller error in variational energy is not necessarily
implying a smaller structural deviation of 2-RDM. The structural deviation
may lead to erroneous prediction of important electronic structure
properties such as the order parameters in condensed state physics.

Explicit violation of inequality constraints by variation 2-RDMs is
not found for the Hubbard model and LiH, most likely, due to the insufficiency
of the universal upper bounds used in our tests. The sharp upper bounds
proposed by Van Neck, Johnson and their coworkers\citep{Neck2007,Johnson2013}
may be useful to ehnace the inequality constraints.

Substituting an eigenoperator of 1-RDM and an eigenoperator of 2-RDM
in Eq.(\ref{eq:ineq_const}), we can also obtain more constraints
on the eigenspace of 1-RDM and 2-RDM. Eq.(\ref{eq:ineq_const}) is
general for operators defined on any Hilbert space. Therefore, the
approach presented here is applicable to characterize not only $N$-representability
of fermions but also that of bosons and quantum marginal problem in
general.

In this work, we derive a set of necessary $N$-representability conditions
on 2-RDMs based on the basic geometric property of Hilbert and the
commutation relations of operators. We show that the algebra properties
of the eigenoperators of 2-RDMs lead to equality constraints on the
null spaces of 2-RDMs. Using triangle inequality, a further analysis
results in a set of inequalities expanding the constraints to the
whole eigenspace of 2-RDMs. Numerical tests show that, compared to
the available positive semidefinite conditions on 2-RDMs, these conditions
impose more stringent constraint on the structure of 2-RDMs. 

Implementing the geometric constraints in ground-state-energy optimization
will not be straightforward due to their nonlinear nature. Incorporation
into SDP may be carried out in a self-consistent manor, in which these
conditions provide correction to varitional RDMs and new approximated
constraints for the next round SDP optimization. However, this may
lead to significant increase of computational cost. Another interesting
direction to explore is their application in hybrid quantum-classical
computing.\citep{Rubin2018,Smart2019a} Recent progresses show that
$N$-representability conditions can be utilized to mitigate quantum
error in electronic structure simulations, and to reduce the number
of required quantum measurements by one order.
\begin{acknowledgments}
The author would like to acknowledge Dr. Maho Nakata at RIKEN and
Dr. Nicholas Robin at Google Research for their valuable comments
and suggestions on the manuscript.
\end{acknowledgments}

\setcounter{table}{0}

\renewcommand{\thetable}{A\arabic{table}}

\setcounter{equation}{0}

\renewcommand{\theequation}{A\arabic{equation}}

\section*{References}

\bibliographystyle{apsrev}

\section*{Appendix}

\subsection*{Derivation of the commutation relations for 2-RDM eigenoperators}

(a) For $g_{m}$ and $g_{n}$, two eigenoperators of $G$ matrix,\begin{subequations}
\begin{align}
\left[g_{m},g_{n}\right] & =\sum_{i,j=1}^{L}\sum_{k,l=1}^{L}\left[v_{ij}^{m}a_{j}^{\dagger}a_{i},v_{kl}^{n}a_{l}^{\dagger}a_{k}\right]\nonumber \\
 & =\sum_{i,j=1}^{L}\sum_{k,l=1}^{L}v_{ij}^{m}v_{kl}^{n}\left[a_{j}^{\dagger}a_{i},a_{l}^{\dagger}a_{k}\right]\nonumber \\
 & =\sum_{i,j=1}^{L}\sum_{k,l=1}^{L}v_{ij}^{m}v_{kl}^{n}\left(a_{j}^{\dagger}a_{k}\delta_{il}-a_{l}^{\dagger}a_{i}\delta_{jk}\right)\nonumber \\
 & =\sum_{j,k,l=1}^{L}v_{lj}^{m}v_{kl}^{n}a_{j}^{\dagger}a_{k}-\sum_{i,k,l=1}^{L}v_{ik}^{m}v_{kl}^{n}a_{l}^{\dagger}a_{i}\nonumber \\
 & =\sum_{i,j=1}^{L}\left[\sum_{k=1}^{L}\left(v_{kj}^{m}v_{ik}^{n}-v_{ik}^{m}v_{kj}^{n}\right)\right]a_{j}^{\dagger}a_{i}\nonumber \\
 & =\sum_{i,j=1}^{L}\overline{v}_{ij}^{mn}a_{j}^{\dagger}a_{i},\label{eq: com_gg1}
\end{align}
 here 
\begin{equation}
\overline{v}_{ij}^{mn}=\sum_{k=1}^{L}\left(v_{kj}^{m}v_{ik}^{n}-v_{ik}^{m}v_{kj}^{n}\right).\label{eq:com_gg2}
\end{equation}

Now expanding $\overline{\boldsymbol{v}}^{mn}$ in the basis set $\left\{ v_{ij}^{m'},m'=1,2,\cdots,L^{2}\right\} $,
we have 
\begin{align}
\left[g_{m},g_{n}\right] & =\sum_{m'=1}^{L^{2}}\sum_{i,j=1}^{L}\left(\sum_{k,l=1}^{L}\overline{v}_{kl}^{mn}v_{kl}^{m'*}\right)v_{ij}^{m'}a_{j}^{\dagger}a_{i}\nonumber \\
 & =\sum_{m'=1}^{L^{2}}\Gamma_{mn}^{m'}g_{m'},\label{eq:com_gg3}
\end{align}
 here

\begin{align}
\Gamma_{mn}^{m'} & =\sum_{i,j=1}^{L}\overline{v}_{ij}^{mn}v_{ij}^{m'*}\nonumber \\
 & =\sum_{i,j,k=1}^{L}\left(v_{kj}^{m}v_{ik}^{n}-v_{ik}^{m}v_{kj}^{n}\right)v_{ij}^{m'*}\label{eq:com_gg4}
\end{align}
\end{subequations}(b) For $g_{m}$ and $d_{n}$, two eigenoperators
of $G$ and $D$ matrix, respectively, \begin{subequations}
\begin{align}
\left[g_{m},d_{n}\right] & =\sum_{i,j=1}^{L}\sum_{k,l=1}^{L}\left[v_{ij}^{m}a_{j}^{\dagger}a_{i},u_{kl}^{n}a_{l}a_{k}\right]\nonumber \\
 & =\sum_{i,j=1}^{L}\sum_{k,l=1}^{L}v_{ij}^{m}u_{kl}^{n}\left[a_{j}^{\dagger}a_{i},a_{l}a_{k}\right]\nonumber \\
 & =\sum_{i,j=1}^{L}\sum_{k,l=1}^{L}v_{ij}^{m}u_{kl}^{n}\left(a_{k}a_{i}\delta_{jl}-a_{l}a_{i}\delta_{jk}\right)\nonumber \\
 & =\sum_{i,k,l=1}^{L}\left(v_{il}^{m}u_{kl}^{n}a_{k}a_{i}-v_{ik}^{m}u_{kl}^{n}a_{l}a_{i}\right)\nonumber \\
 & =\sum_{i,j=1}^{L}\left[\sum_{k=1}^{L}\left(v_{ik}^{m}u_{jk}^{n}-v_{ik}^{m}u_{kj}^{n}\right)\right]a_{j}a_{i}\nonumber \\
 & =\sum_{i,j=1}^{L}\overline{u}_{ij}^{mn}a_{j}a_{i},\label{eq:com_gd1}
\end{align}
 here
\begin{align}
\overline{u}_{ij}^{mn} & =\sum_{k=1}^{L}\left(v_{ik}^{m}u_{jk}^{n}-v_{ik}^{m}u_{kj}^{n}\right)\nonumber \\
 & =-2\sum_{k=1}^{L}v_{ik}^{m}u_{kj}^{n}.\label{eq:com_gd2}
\end{align}
 Here, we have used the fact, $u_{jk}^{n}=-u_{kj}^{n}$. Now expanding
$\overline{\boldsymbol{u}}^{mn}$ in the basis set $\left\{ u_{ij}^{m'},m'=1,2,\cdots,\frac{L\left(L-1\right)}{2}\right\} $,
we have 
\begin{align}
\left[g_{m},d_{n}\right] & =\sum_{m'=1}^{\frac{L\left(L-1\right)}{2}}\sum_{i,j=1}^{L}\left(\sum_{k,l=1}^{L}\overline{u}_{kl}^{mn}u_{kl}^{m'*}\right)u_{ij}^{m'}a_{j}a_{i}\nonumber \\
 & =\sum_{m'=1}^{\frac{L\left(L-1\right)}{2}}\Delta_{mn}^{m'}d_{m'},\label{eq:com_gd3}
\end{align}
 here

\begin{align}
\Delta_{mn}^{m'} & =\sum_{i,j=1}^{L}\overline{u}_{ij}^{mn}u_{ij}^{m'*}\nonumber \\
 & =-2\sum_{i,j,k=1}^{L}v_{ik}^{m}u_{kj}^{n}u_{ij}^{m'*}\label{eq:com_gd4}
\end{align}
\end{subequations}(c) For $g_{m}$ and $q_{n}$, two eigenoperators
of $G$ and $Q$ matrix, respectively, \begin{subequations}
\begin{align}
\left[g_{m},q_{n}\right] & =\sum_{i,j=1}^{L}\sum_{k,l=1}^{L}\left[v_{ij}^{m}a_{j}^{\dagger}a_{i},w_{kl}^{n}a_{l}^{\dagger}a_{k}^{\dagger}\right]\nonumber \\
 & =\sum_{i,j=1}^{L}\sum_{k,l=1}^{L}v_{ij}^{m}w_{kl}^{n}\left[a_{j}^{\dagger}a_{i},a_{l}^{\dagger}a_{k}^{\dagger}\right]\nonumber \\
 & =\sum_{i,j=1}^{L}\sum_{k,l=1}^{L}v_{ij}^{m}w_{kl}^{n}\left(a_{j}^{\dagger}a_{k}^{\dagger}\delta_{il}-a_{j}^{\dagger}a_{l}^{\dagger}\delta_{ik}\right)\nonumber \\
 & =\sum_{j,k,l=1}^{L}\left(v_{lj}^{m}w_{kl}^{n}a_{j}^{\dagger}a_{k}^{\dagger}-v_{kj}^{m}w_{kl}^{n}a_{j}^{\dagger}a_{l}^{\dagger}\right)\nonumber \\
 & =\sum_{i,j=1}^{L}\left[\sum_{k=1}^{L}\left(v_{kj}^{m}w_{ik}^{n}-v_{kj}^{m}w_{ki}^{n}\right)\right]a_{j}^{\dagger}a_{i}^{\dagger}\nonumber \\
 & =\sum_{i,j=1}^{L}\overline{w}_{ij}^{mn}a_{j}^{\dagger}a_{i}^{\dagger},\label{eq:com_gq1}
\end{align}
 here
\begin{align}
\overline{w}_{ij}^{mn} & =\sum_{k=1}^{L}\left(v_{kj}^{m}w_{ik}^{n}-v_{kj}^{m}w_{ki}^{n}\right)\nonumber \\
 & =2\sum_{k=1}^{L}w_{ik}^{n}v_{kj}^{m}.\label{eq:com_gq2}
\end{align}
 Here, we have used the fact, $w_{jk}^{n}=-w_{kj}^{n}$. Now expanding
$\overline{\boldsymbol{w}}^{mn}$ in the basis set $\left\{ w_{ij}^{m'},m'=1,2,\cdots,\frac{L\left(L-1\right)}{2}\right\} $,
we have 
\begin{align}
\left[g_{m},q_{n}\right] & =\sum_{m'=1}^{\frac{L\left(L-1\right)}{2}}\sum_{i,j=1}^{L}\left(\sum_{k,l=1}^{L}\overline{w}_{kl}^{mn}w_{kl}^{m'*}\right)w_{ij}^{m'}a_{j}^{\dagger}a_{i}^{\dagger}\nonumber \\
 & =\sum_{m'=1}^{\frac{L\left(L-1\right)}{2}}\Omega_{mn}^{m'}q_{m'},\label{eq:com_gq3}
\end{align}
 here

\begin{align}
\Omega_{mn}^{m'} & =\sum_{i,j=1}^{L}\overline{w}_{ij}^{mn}w_{ij}^{m'*}\nonumber \\
 & =2\sum_{i,j,k=1}^{L}w_{ik}^{n}v_{kj}^{m}w_{ij}^{m'*}\label{eq:com_gq4}
\end{align}
\end{subequations}(d) For $q_{m}$ and $d_{n}$, two eigenoperators
of $Q$ and $D$ matrix, respectively, \begin{subequations}
\begin{align}
\left[q_{m},d_{n}\right] & =\sum_{i,j=1}^{L}\sum_{k,l=1}^{L}\left[w_{ij}^{m}a_{j}^{\dagger}a_{i}^{\dagger},u_{kl}^{n}a_{l}a_{k}\right]\nonumber \\
 & =\sum_{i,j=1}^{L}\sum_{k,l=1}^{L}w_{ij}^{m}u_{kl}^{n}\left[a_{j}^{\dagger}a_{i}^{\dagger},a_{l}a_{k}\right]\nonumber \\
 & =\sum_{i,j=1}^{L}\sum_{k,l=1}^{L}w_{ij}^{m}u_{kl}^{n}\left(a_{j}^{\dagger}a_{k}\delta_{il}-a_{j}^{\dagger}a_{l}\delta_{ik}+a_{k}a_{i}^{\dagger}\delta_{jl}-a_{l}a_{i}^{\dagger}\delta_{jk}\right)\nonumber \\
 & =\sum_{i,j=1}^{L}\sum_{k,l=1}^{L}w_{ij}^{m}u_{kl}^{n}\left(a_{j}^{\dagger}a_{k}\delta_{il}-a_{j}^{\dagger}a_{l}\delta_{ik}-a_{i}^{\dagger}a_{k}\delta_{jl}+a_{i}^{\dagger}a_{l}\delta_{jk}+\delta_{ik}\delta_{jl}-\delta_{il}\delta_{jk}\right)\nonumber \\
 & =\sum_{j,k,l=1}^{L}\left(w_{lj}^{m}u_{kl}^{n}a_{j}^{\dagger}a_{k}-w_{kj}^{m}u_{kl}^{n}a_{j}^{\dagger}a_{l}\right)+\sum_{i,k,l=1}^{L}\left(-w_{il}^{m}u_{kl}^{n}a_{i}^{\dagger}a_{k}+w_{ik}^{m}u_{kl}^{n}a_{i}^{\dagger}a_{l}\right)+\sum_{k,l=1}^{L}\left(w_{kl}^{m}u_{kl}^{n}-w_{lk}^{m}u_{kl}^{n}\right)\nonumber \\
 & =\sum_{i,j=1}^{L}\left[\sum_{k=1}^{L}\left(w_{kj}^{m}u_{ik}^{n}-w_{lj}^{m}u_{ki}^{n}-w_{jk}^{m}u_{ik}^{n}+w_{jk}^{m}u_{ki}^{n}\right)\right]a_{j}^{\dagger}a_{i}+\sum_{k,l=1}^{L}\left(w_{kl}^{m}u_{kl}^{n}-w_{lk}^{m}u_{kl}^{n}\right)\nonumber \\
 & =\sum_{i,j=1}^{L}\left[\sum_{k=1}^{L}\left(w_{kj}^{m}u_{ik}^{n}-w_{kj}^{m}u_{ki}^{n}-w_{jk}^{m}u_{ik}^{n}+w_{jk}^{m}u_{ki}^{n}\right)+\frac{1}{N}\sum_{k,l=1}^{L}\left(w_{kl}^{m}u_{kl}^{n}-w_{lk}^{m}u_{kl}^{n}\right)\delta_{ij}\right]a_{j}^{\dagger}a_{i}\nonumber \\
 & =\sum_{i,j=1}^{L}\overline{v}_{ij}^{mn}a_{j}^{\dagger}a_{i},\label{eq:com_qd1}
\end{align}
 here, we have restricted ourselves in $N$-electron Fock space, and
\begin{align}
\overline{v}_{ij}^{mn} & =\sum_{k=1}^{L}\left(w_{kj}^{m}u_{ik}^{n}-w_{kj}^{m}u_{ki}^{n}-w_{jk}^{m}u_{ik}^{n}+w_{jk}^{m}u_{ki}^{n}\right)+\frac{1}{N}\sum_{k,l=1}^{L}\left(w_{kl}^{m}u_{kl}^{n}-w_{lk}^{m}u_{kl}^{n}\right)\nonumber \\
 & =4\left(\sum_{k=1}^{L}u_{ik}^{n}w_{kj}^{m}-\frac{1}{2N}\delta_{ij}\sum_{k,l=1}^{L}u_{lk}^{n}w_{kl}^{m}\right)\label{eq:com_qd2}
\end{align}
 Here, we have used the fact, $u_{jk}^{n}=-u_{kj}^{n}$ and $w_{jk}^{n}=-w_{kj}^{n}$.
Now expanding $\overline{\boldsymbol{v}}^{mn}$ in the basis set $\left\{ v_{ij}^{m'},m'=1,2,\cdots,L^{2}\right\} $,
we have 
\begin{align}
\left[q_{m},d_{n}\right] & =4\sum_{m'=1}^{L^{2}}\sum_{i,j=1}^{L}\left(\sum_{k=1}^{L}u_{ik}^{n}w_{kj}^{m}-\frac{1}{2N}\delta_{ij}\sum_{k,l=1}^{L}u_{lk}^{n}w_{kl}^{m}\right)v_{ij}^{m'}a_{j}^{\dagger}a_{i}\nonumber \\
 & =\sum_{m'=1}^{L^{2}}\Theta_{mn}^{m'}g_{m'},\label{eq:com_qd3}
\end{align}
 here
\begin{align}
\Omega_{mn}^{m'} & =\sum_{i,j=1}^{L}\overline{v}_{ij}^{mn}v_{ij}^{m'*}\nonumber \\
 & =4\sum_{i,j,k=1}^{L}\left(u_{ik}^{n}w_{kj}^{m}-\frac{1}{2N}\delta_{ij}\sum_{l=1}^{L}u_{lk}^{n}w_{kl}^{m}\right)v_{ij}^{m'*}\nonumber \\
 & =4\sum_{i,j,k=1}^{L}\left(u_{ik}^{n}w_{kj}^{m}v_{ij}^{m'*}-\frac{1}{2N}u_{ij}^{n}w_{ji}^{m}v_{kk}^{m'*}\right)\label{eq:com_qd4}
\end{align}

\end{subequations}

\begin{table}[h]
\caption{The ten lowest eigenvalues of the variational and exact RDMs for a
6-site half-filled 1D Hubbard model with $t=1$ and $U=10$. The ground
state energies obtained by the two methods are $E_{exact}=-1.664362733287$
and $E_{var}=-1.695384327725$, respectively. The energy deviation
$\Delta E=-0.031021594438$. These energies are consistent with the
previous studies on this model.\citep{Nakata2008}}

\begin{tabular}{|r@{\extracolsep{0pt}.}l|r@{\extracolsep{0pt}.}l|r@{\extracolsep{0pt}.}l|r@{\extracolsep{0pt}.}l|r@{\extracolsep{0pt}.}l|r@{\extracolsep{0pt}.}l|r@{\extracolsep{0pt}.}l|}
\hline 
\multicolumn{2}{|c|}{} & \multicolumn{2}{c}{{\footnotesize{}D matrix}} & \multicolumn{2}{c|}{} & \multicolumn{2}{c}{{\footnotesize{}G matrix}} & \multicolumn{2}{c|}{} & \multicolumn{2}{c}{{\footnotesize{}Q matrix}} & \multicolumn{2}{c|}{}\tabularnewline
\multicolumn{2}{|c|}{{\footnotesize{}n}} & \multicolumn{2}{c}{{\footnotesize{}Variational}} & \multicolumn{2}{c|}{\multirow{1}{*}{{\footnotesize{}Exact}}} & \multicolumn{2}{c}{{\footnotesize{}Variational}} & \multicolumn{2}{c|}{{\footnotesize{}Exact}} & \multicolumn{2}{c}{{\footnotesize{}Variational}} & \multicolumn{2}{c|}{{\footnotesize{}Exact}}\tabularnewline
\hline 
\multicolumn{2}{|c|}{{\footnotesize{}1}} & {\footnotesize{}-0}&{\footnotesize{}000000000001} & {\footnotesize{}-0}&{\footnotesize{}000000000000} & {\footnotesize{}-0}&{\footnotesize{}000000000000} & {\footnotesize{}-0}&{\footnotesize{}000000000000} & {\footnotesize{}-0}&{\footnotesize{}000000000001} & {\footnotesize{}0}&{\footnotesize{}000000000000}\tabularnewline
\hline 
\multicolumn{2}{|c|}{{\footnotesize{}2}} & {\footnotesize{}-0}&{\footnotesize{}000000000001} & {\footnotesize{}0}&{\footnotesize{}000013764099} & {\footnotesize{}-0}&{\footnotesize{}000000000000} & {\footnotesize{}-0}&{\footnotesize{}000000000000} & {\footnotesize{}-0}&{\footnotesize{}000000000001} & {\footnotesize{}0}&{\footnotesize{}000013764099}\tabularnewline
\hline 
\multicolumn{2}{|c|}{{\footnotesize{}3}} & {\footnotesize{}-0}&{\footnotesize{}000000000001} & {\footnotesize{}0}&{\footnotesize{}000558839696} & {\footnotesize{}-0}&{\footnotesize{}000000000000} & {\footnotesize{}0}&{\footnotesize{}000000000000} & {\footnotesize{}-0}&{\footnotesize{}000000000001} & {\footnotesize{}0}&{\footnotesize{}000558839696}\tabularnewline
\hline 
\multicolumn{2}{|c|}{{\footnotesize{}4}} & {\footnotesize{}-0}&{\footnotesize{}000000000001} & {\footnotesize{}0}&{\footnotesize{}000558839696} & {\footnotesize{}-0}&{\footnotesize{}000000000000} & {\footnotesize{}0}&{\footnotesize{}000006882049} & {\footnotesize{}-0}&{\footnotesize{}000000000001} & {\footnotesize{}0}&{\footnotesize{}000558839696}\tabularnewline
\hline 
\multicolumn{2}{|c|}{{\footnotesize{}5}} & {\footnotesize{}-0}&{\footnotesize{}000000000001} & {\footnotesize{}0}&{\footnotesize{}000649120541} & {\footnotesize{}-0}&{\footnotesize{}000000000000} & {\footnotesize{}0}&{\footnotesize{}000279419848} & {\footnotesize{}-0}&{\footnotesize{}000000000001} & {\footnotesize{}0}&{\footnotesize{}000649120541}\tabularnewline
\hline 
\multicolumn{2}{|c|}{{\footnotesize{}6}} & {\footnotesize{}-0}&{\footnotesize{}000000000000} & {\footnotesize{}0}&{\footnotesize{}000649120541} & {\footnotesize{}0}&{\footnotesize{}000000000000} & {\footnotesize{}0}&{\footnotesize{}000279419848} & {\footnotesize{}-0}&{\footnotesize{}000000000000} & {\footnotesize{}0}&{\footnotesize{}000649120541}\tabularnewline
\hline 
\multicolumn{2}{|c|}{{\footnotesize{}7}} & {\footnotesize{}0}&{\footnotesize{}052344794130} & {\footnotesize{}0}&{\footnotesize{}054771204301} & {\footnotesize{}0}&{\footnotesize{}000000000000} & {\footnotesize{}0}&{\footnotesize{}000324560271} & {\footnotesize{}0}&{\footnotesize{}052344794122} & {\footnotesize{}0}&{\footnotesize{}054771204301}\tabularnewline
\hline 
\multicolumn{2}{|c|}{{\footnotesize{}8}} & {\footnotesize{}0}&{\footnotesize{}052344794130} & {\footnotesize{}0}&{\footnotesize{}054771204301} & {\footnotesize{}0}&{\footnotesize{}000000000000} & {\footnotesize{}0}&{\footnotesize{}000324560271} & {\footnotesize{}0}&{\footnotesize{}052344794122} & {\footnotesize{}0}&{\footnotesize{}054771204301}\tabularnewline
\hline 
\multicolumn{2}{|c|}{{\footnotesize{}9}} & {\footnotesize{}0}&{\footnotesize{}053848732617} & {\footnotesize{}0}&{\footnotesize{}056628436116} & {\footnotesize{}0}&{\footnotesize{}019332198318} & {\footnotesize{}0}&{\footnotesize{}026343666870} & {\footnotesize{}0}&{\footnotesize{}053848732627} & {\footnotesize{}0}&{\footnotesize{}056628436116}\tabularnewline
\hline 
\multicolumn{2}{|c|}{{\footnotesize{}10}} & {\footnotesize{}0}&{\footnotesize{}054329721649} & {\footnotesize{}0}&{\footnotesize{}059289055684} & {\footnotesize{}0}&{\footnotesize{}025473863823} & {\footnotesize{}0}&{\footnotesize{}027040912995} & {\footnotesize{}0}&{\footnotesize{}054329721653} & {\footnotesize{}0}&{\footnotesize{}059289055684}\tabularnewline
\hline 
\end{tabular}
\end{table}

\begin{table}[h]
\caption{The five lowest eigenvalues of the variational and exact RDMs for
diatomic molecule LiH. The ground state energies obtained by the two
methods are $E_{exact}=-8.967211312701\,(E_{h})$ and $E_{var}=-8.967211329766\,(E_{h})$,
respectively. The energy deviation $\Delta E=-1.707\times10^{-8}\,(E_{h})$.
These energies are consistent with the previous studies on this system.\citep{Zhao2004,Nakata2008}}

\begin{tabular}{|r@{\extracolsep{0pt}.}l|r@{\extracolsep{0pt}.}l|r@{\extracolsep{0pt}.}l|r@{\extracolsep{0pt}.}l|r@{\extracolsep{0pt}.}l|r@{\extracolsep{0pt}.}l|r@{\extracolsep{0pt}.}l|}
\hline 
\multicolumn{2}{|c|}{} & \multicolumn{2}{c}{{\footnotesize{}D matrix }} & \multicolumn{2}{c|}{} & \multicolumn{2}{c}{{\footnotesize{}G matrix}} & \multicolumn{2}{c|}{} & \multicolumn{2}{c}{{\footnotesize{}Q matrix}} & \multicolumn{2}{c|}{}\tabularnewline
\multicolumn{2}{|c|}{{\footnotesize{}n}} & \multicolumn{2}{c}{{\footnotesize{}Variational }} & \multicolumn{2}{c|}{\multirow{1}{*}{{\footnotesize{}Exact }}} & \multicolumn{2}{c}{{\footnotesize{}Variational }} & \multicolumn{2}{c|}{{\footnotesize{}Exact}} & \multicolumn{2}{c}{{\footnotesize{}Variational}} & \multicolumn{2}{c|}{{\footnotesize{}Exact}}\tabularnewline
\hline 
\multicolumn{2}{|c|}{{\footnotesize{}1}} & {\footnotesize{}0}&{\footnotesize{}000000000671 } & {\footnotesize{}0}&{\footnotesize{}000000001019 } & {\footnotesize{}-0}&{\footnotesize{}000000000000 } & {\footnotesize{}-0}&{\footnotesize{}000000000000 } & {\footnotesize{}0}&{\footnotesize{}000000000137 } & {\footnotesize{}0}&{\footnotesize{}000000014488073 }\tabularnewline
\hline 
\multicolumn{2}{|c|}{{\footnotesize{}2}} & {\footnotesize{}0}&{\footnotesize{}000000000920} & {\footnotesize{}0}&{\footnotesize{}000000011154} & {\footnotesize{}0}&{\footnotesize{}000000000000} & {\footnotesize{}0}&{\footnotesize{}000000000000} & {\footnotesize{}0}&{\footnotesize{}000001131230} & {\footnotesize{}0}&{\footnotesize{}000001027814592}\tabularnewline
\hline 
\multicolumn{2}{|c|}{{\footnotesize{}3}} & {\footnotesize{}0}&{\footnotesize{}000000004001} & {\footnotesize{}0}&{\footnotesize{}000000011154} & {\footnotesize{}-0}&{\footnotesize{}000000000000} & {\footnotesize{}-0}&{\footnotesize{}000000000000} & {\footnotesize{}0}&{\footnotesize{}000001245966} & {\footnotesize{}0}&{\footnotesize{}000001179234760}\tabularnewline
\hline 
\multicolumn{2}{|c|}{{\footnotesize{}4}} & {\footnotesize{}0}&{\footnotesize{}000000004001} & {\footnotesize{}0}&{\footnotesize{}000000011154} & {\footnotesize{}-0}&{\footnotesize{}000000000000} & {\footnotesize{}-0}&{\footnotesize{}000000000000} & {\footnotesize{}0}&{\footnotesize{}000001245966} & {\footnotesize{}0}&{\footnotesize{}000001179234760}\tabularnewline
\hline 
\multicolumn{2}{|c|}{{\footnotesize{}5}} & {\footnotesize{}0}&{\footnotesize{}000000007047} & {\footnotesize{}0}&{\footnotesize{}000000011154} & {\footnotesize{}-0}&{\footnotesize{}000000000863} & {\footnotesize{}-0}&{\footnotesize{}000000000000} & {\footnotesize{}0}&{\footnotesize{}000001245966} & {\footnotesize{}0}&{\footnotesize{}000001179234760}\tabularnewline
\hline 
\end{tabular}
\end{table}

\begin{table}[h]
\caption{The eight lowest eigenvalues of the variational and exact RDMs for
random hamiltonian 1. The hamiltonian matrix elements in the spatial
degree of freedom are randomly generated with a uniform distribution
in $[0,1)$.The ground state energies obtained by the two methods
are $E_{exact}=-5.474591092179$ and $E_{var}=-5.601757205138$, respectively.
The energy deviation $\Delta E=-0.127166112960$. }

\begin{tabular}{|r@{\extracolsep{0pt}.}l|r@{\extracolsep{0pt}.}l|r@{\extracolsep{0pt}.}l|r@{\extracolsep{0pt}.}l|r@{\extracolsep{0pt}.}l|r@{\extracolsep{0pt}.}l|r@{\extracolsep{0pt}.}l|}
\hline 
\multicolumn{2}{|c|}{} & \multicolumn{2}{c}{{\footnotesize{}D matrix }} & \multicolumn{2}{c|}{} & \multicolumn{2}{c}{{\footnotesize{}G matrix}} & \multicolumn{2}{c|}{} & \multicolumn{2}{c}{{\footnotesize{}Q matrix}} & \multicolumn{2}{c|}{}\tabularnewline
\multicolumn{2}{|c|}{{\footnotesize{}n}} & \multicolumn{2}{c}{{\footnotesize{}Variational }} & \multicolumn{2}{c|}{\multirow{1}{*}{{\footnotesize{}Exact }}} & \multicolumn{2}{c}{{\footnotesize{}Variational }} & \multicolumn{2}{c|}{{\footnotesize{}Exact}} & \multicolumn{2}{c}{{\footnotesize{}Variational}} & \multicolumn{2}{c|}{{\footnotesize{}Exact}}\tabularnewline
\hline 
\multicolumn{2}{|c|}{{\footnotesize{}1}} & {\footnotesize{}0}&{\footnotesize{}000000000000 } & {\footnotesize{}0}&{\footnotesize{}000143928633} & {\footnotesize{}-0}&{\footnotesize{}000000000000 } & {\footnotesize{}-0}&{\footnotesize{}000000000000 } & {\footnotesize{}0}&{\footnotesize{}000000000017 } & {\footnotesize{}0}&{\footnotesize{}001790119018 }\tabularnewline
\hline 
\multicolumn{2}{|c|}{{\footnotesize{}2}} & {\footnotesize{}0}&{\footnotesize{}000613559566} & {\footnotesize{}0}&{\footnotesize{}001929013145} & {\footnotesize{}-0}&{\footnotesize{}000000000000} & {\footnotesize{}-0}&{\footnotesize{}000000000000} & {\footnotesize{}0}&{\footnotesize{}000000000017} & {\footnotesize{}0}&{\footnotesize{}002205303031}\tabularnewline
\hline 
\multicolumn{2}{|c|}{{\footnotesize{}3}} & {\footnotesize{}0}&{\footnotesize{}000613559566} & {\footnotesize{}0}&{\footnotesize{}001929013145} & {\footnotesize{}-0}&{\footnotesize{}000000000000} & {\footnotesize{}0}&{\footnotesize{}000000000000} & {\footnotesize{}0}&{\footnotesize{}000000000017} & {\footnotesize{}0}&{\footnotesize{}002205303031}\tabularnewline
\hline 
\multicolumn{2}{|c|}{{\footnotesize{}4}} & {\footnotesize{}0}&{\footnotesize{}000613559566} & {\footnotesize{}0}&{\footnotesize{}001929013145} & {\footnotesize{}0}&{\footnotesize{}000000000006} & {\footnotesize{}0}&{\footnotesize{}000653753069} & {\footnotesize{}0}&{\footnotesize{}005038116249} & {\footnotesize{}0}&{\footnotesize{}002205303031}\tabularnewline
\hline 
\multicolumn{2}{|c|}{{\footnotesize{}5}} & {\footnotesize{}0}&{\footnotesize{}005911622327} & {\footnotesize{}0}&{\footnotesize{}002005631157} & {\footnotesize{}0}&{\footnotesize{}000000000006} & {\footnotesize{}0}&{\footnotesize{}000653753069} & {\footnotesize{}0}&{\footnotesize{}007038945113} & {\footnotesize{}0}&{\footnotesize{}004683583778}\tabularnewline
\hline 
\multicolumn{2}{|c|}{{\footnotesize{}6}} & {\footnotesize{}0}&{\footnotesize{}005911622327} & {\footnotesize{}0}&{\footnotesize{}003709814514} & {\footnotesize{}0}&{\footnotesize{}000000000006} & {\footnotesize{}0}&{\footnotesize{}000653753069} & {\footnotesize{}0}&{\footnotesize{}007038945113} & {\footnotesize{}0}&{\footnotesize{}004770757563}\tabularnewline
\hline 
\multicolumn{2}{|c|}{{\footnotesize{}7}} & {\footnotesize{}0}&{\footnotesize{}005911622327} & {\footnotesize{}0}&{\footnotesize{}003709814514} & {\footnotesize{}0}&{\footnotesize{}000000000006} & {\footnotesize{}0}&{\footnotesize{}000821845808} & {\footnotesize{}0}&{\footnotesize{}007038945113} & {\footnotesize{}0}&{\footnotesize{}004770757563}\tabularnewline
\hline 
\multicolumn{2}{|c|}{{\footnotesize{}8}} & {\footnotesize{}0}&{\footnotesize{}006553183666} & {\footnotesize{}0}&{\footnotesize{}003709814514} & {\footnotesize{}0}&{\footnotesize{}001077905210} & {\footnotesize{}0}&{\footnotesize{}000908689956} & {\footnotesize{}0}&{\footnotesize{}017400000968} & {\footnotesize{}0}&{\footnotesize{}004770757563}\tabularnewline
\hline 
\end{tabular}
\end{table}

\begin{table}[h]
\caption{The eight lowest eigenvalues of the variational and exact RDMs for
random Hamiltonian 2. The Hamiltonian matrix elements in the spatial
degree of freedom are randomly generated with a uniform distribution
in $[0,1)$.The ground state energies obtained by the two methods
are $E_{exact}=-9.559169540991$ and $E_{var}=-9.571940687877$, respectively.
The energy deviation $\Delta E=-0.012771146886$.}

\begin{tabular}{|r@{\extracolsep{0pt}.}l|r@{\extracolsep{0pt}.}l|r@{\extracolsep{0pt}.}l|r@{\extracolsep{0pt}.}l|r@{\extracolsep{0pt}.}l|r@{\extracolsep{0pt}.}l|r@{\extracolsep{0pt}.}l|}
\hline 
\multicolumn{2}{|c|}{} & \multicolumn{2}{c}{{\footnotesize{}D matrix }} & \multicolumn{2}{c|}{} & \multicolumn{2}{c}{{\footnotesize{}G matrix}} & \multicolumn{2}{c|}{} & \multicolumn{2}{c}{{\footnotesize{}Q matrix}} & \multicolumn{2}{c|}{}\tabularnewline
\multicolumn{2}{|c|}{{\footnotesize{}n}} & \multicolumn{2}{c}{{\footnotesize{}Variational }} & \multicolumn{2}{c|}{\multirow{1}{*}{{\footnotesize{}Exact }}} & \multicolumn{2}{c}{{\footnotesize{}Variational }} & \multicolumn{2}{c|}{{\footnotesize{}Exact}} & \multicolumn{2}{c}{{\footnotesize{}Variational}} & \multicolumn{2}{c|}{{\footnotesize{}Exact}}\tabularnewline
\hline 
\multicolumn{2}{|c|}{{\footnotesize{}1}} & {\footnotesize{}0}&{\footnotesize{}000000000005 } & {\footnotesize{}0}&{\footnotesize{}000021168521} & {\footnotesize{}-0}&{\footnotesize{}000000000000} & {\footnotesize{}-0}&{\footnotesize{}000000000000 } & {\footnotesize{}0}&{\footnotesize{}000000000221 } & {\footnotesize{}0}&{\footnotesize{}000127020266940 }\tabularnewline
\hline 
\multicolumn{2}{|c|}{{\footnotesize{}2}} & {\footnotesize{}0}&{\footnotesize{}000000000716} & {\footnotesize{}0}&{\footnotesize{}000066283360} & {\footnotesize{}-0}&{\footnotesize{}000000000000} & {\footnotesize{}0}&{\footnotesize{}000000000000} & {\footnotesize{}0}&{\footnotesize{}000700446139} & {\footnotesize{}0}&{\footnotesize{}000127020266941}\tabularnewline
\hline 
\multicolumn{2}{|c|}{{\footnotesize{}3}} & {\footnotesize{}0}&{\footnotesize{}000000000716} & {\footnotesize{}0}&{\footnotesize{}000066283360} & {\footnotesize{}-0}&{\footnotesize{}000000000000} & {\footnotesize{}0}&{\footnotesize{}000000000000} & {\footnotesize{}0}&{\footnotesize{}000700446139} & {\footnotesize{}0}&{\footnotesize{}000127020266941}\tabularnewline
\hline 
\multicolumn{2}{|c|}{{\footnotesize{}4}} & {\footnotesize{}0}&{\footnotesize{}000000000716} & {\footnotesize{}0}&{\footnotesize{}000066283360} & {\footnotesize{}0}&{\footnotesize{}000000000045} & {\footnotesize{}0}&{\footnotesize{}000029660999} & {\footnotesize{}0}&{\footnotesize{}000700446139} & {\footnotesize{}0}&{\footnotesize{}000139446906045}\tabularnewline
\hline 
\multicolumn{2}{|c|}{{\footnotesize{}5}} & {\footnotesize{}0}&{\footnotesize{}001002468131} & {\footnotesize{}0}&{\footnotesize{}000151406632} & {\footnotesize{}0}&{\footnotesize{}000000000052} & {\footnotesize{}0}&{\footnotesize{}000029660999} & {\footnotesize{}0}&{\footnotesize{}000896131590} & {\footnotesize{}0}&{\footnotesize{}000198318040334}\tabularnewline
\hline 
\multicolumn{2}{|c|}{{\footnotesize{}6}} & {\footnotesize{}0}&{\footnotesize{}001135685285} & {\footnotesize{}0}&{\footnotesize{}000178358468} & {\footnotesize{}0}&{\footnotesize{}000000000052} & {\footnotesize{}0}&{\footnotesize{}000029660999} & {\footnotesize{}0}&{\footnotesize{}000935564244} & {\footnotesize{}0}&{\footnotesize{}000198318040334}\tabularnewline
\hline 
\multicolumn{2}{|c|}{{\footnotesize{}7}} & {\footnotesize{}0}&{\footnotesize{}001135685285} & {\footnotesize{}0}&{\footnotesize{}000178358468} & {\footnotesize{}0}&{\footnotesize{}000000000052} & {\footnotesize{}0}&{\footnotesize{}000037531996} & {\footnotesize{}0}&{\footnotesize{}000935564244} & {\footnotesize{}0}&{\footnotesize{}000198318040334}\tabularnewline
\hline 
\multicolumn{2}{|c|}{{\footnotesize{}8}} & {\footnotesize{}0}&{\footnotesize{}001135685285} & {\footnotesize{}0}&{\footnotesize{}000178358468} & {\footnotesize{}0}&{\footnotesize{}000149458470} & {\footnotesize{}0}&{\footnotesize{}000043974367} & {\footnotesize{}0}&{\footnotesize{}000935564244} & {\footnotesize{}0}&{\footnotesize{}000218197581049}\tabularnewline
\hline 
\end{tabular}
\end{table}

\end{document}